\documentclass[final,leqno,onefignum,onetabnum]{siamltex1213}

\usepackage{amsmath,amssymb}

\usepackage{bbm}
\usepackage{color}

\newtheorem{remark}[theorem]{Remark}
\newtheorem{algorithm}[theorem]{Algorithm}

\newcommand{\pd}{\mathrm{d}}

\newcommand{\Prob}{\mathbb{P}}

\newcommand{\Om}{\mathcal{A}_{\mathrm{m}}}

\newcommand{\Pm}{P_{\mathrm{m}}}
\newcommand{\Pms}{\pi_{\mathrm{m}}}

\newcommand{\Pmsse}{\Pi_{\mathrm{m}}}

\newcommand{\Of}{\mathcal{A}_{\mathrm{f}}}

\newcommand{\Ofb}{\mathcal{A}_\Omega}
\newcommand{\Pf}{P_{\mathrm{f}}}
\newcommand{\Pfs}{\pi_{\mathrm{f}}}

\newcommand{\Pfsse}{\Pi_{\mathrm{f}}}

\newcommand{\Pfb}{P_{\Omega}} 

\newcommand{\Cdif}{a}
\newcommand{\Cdrif}{b}

\newcommand{\alp}{\alpha}

\newcommand{\rate}{k}

\newcommand{\Os}{E} 
\newcommand{\Vs}{\nu}  

\newcommand{\Ns}{N} 
\newcommand{\Nr}{M} 

\newcommand{\Vol}{V} 

\newcommand{\D}{\Omega} 

\newcommand{\gridh}{h} 

\newcommand{\shift}{\sigma}

\newcommand{\bfx}{{\boldsymbol{x}}}

\newcommand{\bfn}{{\boldsymbol{n}}}
\newcommand{\bfp}{{\boldsymbol{p}}}
\newcommand{\bfphi}{{\boldsymbol{\phi}}}
\newcommand{\bfvarepsilon}{{\boldsymbol{\varepsilon}}}

\title{Analysis of tensor methods for stochastic models of gene regulatory networks} 

\author{Shuohao Liao\thanks{e-mails: shuohao.liao@gmail.com}}

\begin{document}
\maketitle
\slugger{mms}{xxxx}{xx}{x}{x--x}

\begin{abstract}
The tensor-structured parametric analysis (TPA) has been recently developed for simulating and analysing stochastic behaviours of gene regulatory networks [Liao et. al., 2015].
The method employs the Fokker-Planck approximation of the chemical master equation, and uses the Quantized Tensor Train (QTT) format, as a low-parametric tensor-structured representation of classical matrices and vectors, to approximate the high-dimensional stationary probability distribution.
This paper presents a detailed error analysis of all approximation steps of the TPA regarding validity and accuracy, including modelling error, artificial boundary error, discretization error, tensor rounding error, and algebraic error.
The error analysis is illustrated using computational examples, including the death-birth process and  a 50-dimensional isomerization reaction chain.
\end{abstract}



\pagestyle{myheadings}
\thispagestyle{plain}
\markboth{S. LIAO}{TENSOR SIMULATION METHODS}

\section{Introduction}

\subsection{Stochastic modelling}

We briefly review the two widely-used mathematical formulations of stochastic reaction networks.
A well-mixed chemically reacting system of $N$ distinct molecular species inside a reactor of volume $\Vol$ is described, at time $t$, by its $N$-dimensional state vector
\begin{align*}
	\mathbf{X}(t) \equiv [X_1(t), X_2(t), \ldots, X_N (t)]^T,
\end{align*}
where $X_i(t) = x_i$ is the number of molecules of the $i$-th chemical species.
 We assume that molecules interact through $\Nr$ reaction channels
\begin{align} \label{eq: mass-action reactions}
\sum_{i=1}^\Ns \Vs^-_{j,i}X_i
\overset{\rate_j}{\longrightarrow} 
\sum_{i=1}^\Ns \Vs^+_{j,i}X_i,
\quad j=1,2,\dots,\Nr,
\end{align}
where $\Vs^+_{j,i}$ and $\Vs^-_{j,i}$ are the stoichiometric coefficients. 
The kinetic rate parameters, $\mathbf{\rate} = (\rate_1,\rate_2,\ldots, \rate_\Nr)^T$, characterise the speed of the corresponding chemical reactions. 
Let $\Pms (\bfn) \equiv \Prob \{\lim_{t\rightarrow\infty} \mathbf{X}(t) = \bfn\}$ be the probability of the system being in the state $\bfn \in \mathbb{N}^N$ in the steady state.
The exact description of such probability is the stationary chemical master equation (CME) of the form
\begin{align} \label{eq: CME}
\Om (\bfn) \Pms (\bfn) \equiv 
	\sum_{j=1}^M \left( \Os^{-\boldsymbol{\Vs}_{j}} - 1 \right) [ \alp_j (\bfn) \Pms \left( \bfn \right)] =0,
	\quad \bfn \in \mathbb{N}^N,
\end{align}
where $\Om (\bfn) $ denotes the CME operator, and $\Os^{-\boldsymbol{\Vs}_j}$ represents the step operator that replaces the arguments of some function $\bfn$ by $\bfn-\boldsymbol{\Vs}_j$, i.e., $\Os^{-\boldsymbol{\Vs}_j} f(\bfn) = f(\bfn - \boldsymbol{\Vs}_j)$.
We denote $\boldsymbol{\Vs}_j = [\Vs_{j,1},\Vs_{j,2},\ldots,\Vs_{j,\Ns}]^T$ by the $j$-th column of the stoichiometric matrix, $[\Vs_{j,i}]_{M\times N}$, with $\Vs_{j,i} = \Vs_{j,i}^+ - \Vs_{j,i}^-$. 
The propensity function $\alp_j$ is an interpretation of the occurrence tendency of the $j$-th reaction. 
Using the mass action kinetics, the propensity functions are of the form
\begin{align*}
	\alp_j (\bfx) = 
	k_j 
	\Vol^{1-o_j} 
	\prod_{i=1}^N 
	\beta_{j,i}(x_i),
	\quad
	\beta_{j,i}(x_i) = \left\{ 
	  \begin{array}{ll}
	   \displaystyle \frac{x_i (x_i-1) \dots (x_i - \Vs_{j,i}^- + 1)}{ (\Vs_{j,i}^-) !} & \text{if } \Vs_{j,i}^- \neq 0, \\
	   1 & \text{if } \Vs_{j,i}^- = 0,
	  \end{array}
	\right.
\end{align*}
for $\bfx \in [0,\infty)^N$,
where $o_j = \sum_{i=1}^\Ns \Vs_{j,i}^-$ stands for the order of the reaction $j = 1,2,\dots,M$.
Since the elementary reactions often involve at most interactions of two molecules, it is often assumed that $o_j \leq 2$ for all $j = 1, 2, \ldots, M$.

Seeking $\Pms$ satisfying \eqref{eq: CME} is the problem of finding the kernel of the linear operator $\Om$. It can be interpreted as finding the eigenfunction $\Pms$ corresponding to the zero eigenvalue of $\Om$.
It can be shown that the CME \eqref{eq: CME} has a unique solution satisfying the natural normalization condition
$
  \sum_{\bfn \in \mathbb{N}^N} \Pms(\bfn) = 1.
$
Solving this exactly is intractable in general, however, it is possible to truncate the positive orthant by imposing a sufficiently large maximum copy number of each chemical species based on the reachability of states~\cite{munsky2006the}.
Then the solution $\Pms$ is approximately computed as the principal eigenfunction of a finite dimensional truncation of $\Om$.

One of the main disadvantages of the CME \eqref{eq: CME} is its analytical intractability for reaction systems which involve higher-order reactions.
Kramers~\cite{kramers1940brownian} and Moyal~\cite{moyal1949the,gillespie2000the} derived a continuous approximation of the stationary CME as a linear, diffusion-convection partial differential equation as
\begin{align} \label{eq: CFPE}
	\Of (\bfx) \Pfs (\bfx)
	= \sum_{i,j=1}^N \partial_{i,j} [a_{i,j} \Pf (\bfx)]
	+ \sum_{i=1}^N \partial_i [b_i \Pfs (\bfx)] 
	= 0,
	\quad
	\bfx \in \Omega_\infty,
\end{align}
where the diffusion and drift coefficients are respectively given by
\begin{align} \label{eq: continuous FP eigenvalue problem diffusion drift coefficients}
\Cdif_{i,j} (\bfx) = \frac{1}{2} \sum_{r=1}^M \Vs_{r,j} \Vs_{r,i} \alp_{r} (\bfx)
	,
	\ \ 
	\Cdrif_j (\bfx) = -\sum_{r=1}^M \Vs_{r,j} \alp_r (\bfx),
	\ \ 
	\mathrm{for}
	\ 
	i,j = 1, 2, \ldots, N.
\end{align}
The domain of definition, denoted as $\Omega_\infty$, is an open domain such that the ellipticity condition is satisfied, i.e.,
\begin{align} \label{eq: ellitic conditions in Omega^infty}
	\Omega_\infty = \left\{ \bfx \in \mathbb{R}^{N}: 
	\exists c_1, c_2 > 0, 
	\forall \boldsymbol{\xi} \in \mathbb{R}^{N},
	c_1 \sum_{i=1}^N \xi_i^2 \leq \sum_{i,j=1}^N a_{i,j} (\bfx) \xi_i \xi_j \leq c_2 \sum_{i=1}^N \xi_i^2 \right\}.
\end{align}
Eq. \eqref{eq: CFPE} is known as the stationary chemical Fokker-Planck equation (CFPE). 
It describes how likely the system will be in a certain portion of the state space, rather than a particular integer state in \eqref{eq: CME}.
So a major and important difference between the CME and the CFPE is that $x_i$ is a non-negative integer for the CME and a real number for the CFPE.

The question of boundary conditions for the CFPE \eqref{eq: continuous FP eigenvalue problem diffusion drift coefficients} is delicate, because unlike the CME, which guarantees that the support of the probability density $\Pms$ lies within the positive orthant, the CFPE can give rise to negative copy-numbers of the chemical species. 
We will consider $\Pfs$ to be a function defined in $\Omega_\infty$ as a solution of \eqref{eq: CFPE}, and satisfy the nomalisation condition $\int_{\Omega_\infty} \Pfs (\bfx) \pd \bfx = 1$.
It can be shown that, for sufficiently large system volume $\Vol$, the stationary distribution $\Pfs$ is bounded and unique in $\Omega_\infty$, with vanishing values along the boundary $\partial \Omega_\infty$.
We refer the readers to \cite{liao2016high} for details on this issue.

Analytical solutions of CFPE remain elusive for many complex biological systems, and numerical simulation techniques are essential in practical applications.

\subsection{Tensor formalism}

A fundamental difficulty of the traditional approaches to solve the CME \eqref{eq: CME} and the CFPE \eqref{eq: CFPE} is the so-called \emph{curse of dimensionality}~\cite{bellman1961adaptive}. 
It refers a universal feature of classic matrix-vector-based data format that the memory requirements and computational complexity of basic arithmetic operations grow exponentially in the number of dimensions, $N$.
As a consequence, both equations \eqref{eq: CME} and \eqref{eq: CFPE} have been historically simulated using the kinetic Monte Carlo methods, such as the Gillespie stochastic simulation algorithm (SSA)  \cite{gillespie1977exact} and its equivalent formulations \cite{gibson2000efficient,cao2004the}.
These approaches generate statistically correct trajectories, and sample probability distributions. 
A disadvantage is that they require many realizations to sample in the very low probability regions, or rare events.

Tensor representation has recently been developed to address the ``curse of dimensionality''~\cite{kolda2009tensor}.
Tensors are multidimensional arrays of real numbers, upon which algebraic operations generalizing matrix-vector-based operations can be performed.
Through generalizing the singular value decomposition (SVD) for matrices to tensors, one could obtains various low-parametric representations of tensors, such as canonical polyadic (CP) representation~\cite{hitchcock1927the}, Tucker representation~\cite{tucker1966some}, tensor train (TT) representation~\cite{oseledets2011tensor}, and hierarchical Tucker representation~\cite{grasedyck2010hierarchical}. 
Recently, the time-dependent version of the CME \eqref{eq: CME} has been re-formualted into tensor formats, and solved directly using the time-stepping procedures under the tensor framework~\cite{jahnke2008a,dolgove2014simultaneous,kazeev2014direct}.

In \cite{liao2014parameter}, the authors introduced the tensor parametric analysis (TPA) and solved the high-dimensional stationary CFPE \eqref{eq: CFPE} using the recently proposed Quantized Tensor Train (QTT) format. 
Table \ref{tab: TPA} demonstrates the simulation steps of the TPA method.
Each step introduces a different type of approximation, and contributes an additional error to the resulting approximate solution of the stationary CME \eqref{eq: CME}.
Therefore, in this paper, we undertake a detailed analysis of the validity of all these approximation steps, and study the convergence of different sources of error that the TPA method incurs.

\begin{table}[t] 
\caption{\label{tab: TPA}
 Schematic procedure of the TPA \cite{liao2014parameter}.
 }
\hrule
\vskip 2mm
\hangindent 8mm {\bf (a1)}  
The solution of the stationary CME \eqref{eq: CME}, $\Pms: \mathbb{N}^N \mapsto \mathbb{R}$, is approximated by the solution of the stationary CFPE \eqref{eq: CFPE}, $\Pfs: \Omega_\infty \mapsto \mathbb{R}$.

\hangindent 8mm {\bf (a2)} 
The CFPE is truncated into a bounded domain $\D \subset \D_\infty $ and $\Pfs$ approximated by $\Pfb$ given by a Dirichlet eigenvalue problem $\D$. 

\hangindent 7.5mm {\bf (a3)} 
The Dirichlet eigenvalue problem in the bounded domain $\D$ is discretised by a finite difference scheme and $\Pfb$ is approximated by a discrete solution $p_h$.

\hangindent 7.5mm {\bf (a4)} 
The discrete problem is solved using the QTT tensor format and the tensor rank of the discrete operator is is truncated. 
Consequently, the discrete solution $p_h$ is approximated by a tensor format solution $\bfp_{\mathrm{t}}$.

\hangindent 7.5mm {\bf (a5)} 
The truncated tensor problem is solved by a tensor-structured iterative method. After $k$ iterations, the algorithm generates a QTT tensor $\bfp_k$ as an approximation of $\bfp_{\mathrm{t}}$.

\vskip 2mm
\hrule
\end{table}

\subsection{Error identification}

Starting from step {\bf (a1)} to {\bf (a5)}, the error in each step is identified.
Detailed mathematical formulations and studies of all these steps are presented separately in Sections 2--6, and numerical verifications are given in Section 7.
We summarise our results below.

\paragraph{Modelling error}

In step {\bf (a1)}, the CME \eqref{eq: CME} is approximated by the CFPE \eqref{eq: CFPE}. 
The approximation is obtained by a perfunctory second-order truncation of the Taylor expansion of the CME (see Section 2.1)~\cite{gillespie2000the}.
A few studies have suggested the CFPE's validity in the thermodynamic limit, where the system volume $V$ approaches to infinity.
Kurtz~\cite{kurtz1978the} proves that the difference between the jump and continuous Markov processes is of order $\log (\Vol) / \Vol$.
Grima \emph{et al.}~\cite{grima2011how} used system-size expansion to show the CFPE predictions of the mean and the variance are accurate to order $\Vol^{-3/2}$.
Here, the tensor methods seek to simulate the whole probability distribution, rather than the summary statistics, the error between the CME and the CFPE distributions are of main interest.
In Section 2, we apply the system-size expansion techniques in \cite{grima2011how} to estimate the $\infty$-norm between $\Pfs$ and $\Pms$. 
In Theorem \ref{theorem: modellingerrgeneral}, we show that the difference is of order $\mathcal{O} \left( \Vol^{-(N+1)/2} \right)$ for general reaction networks.
We also provide a tighter bound as $\mathcal{O} \left( \Vol^{-(N+2)/2} \right)$ in Theorem \ref{theorem: modellingerrdetailbalance} for systems satisfying the detailed balanced condition.

\paragraph{Artificial boundary error}

In step {\bf (a2)}, the TPA solves the Dirichlet eigenvalue problem on a bounded domain $\D \subset \D^\infty$ in which the vast majority of the probability density sits, and then uses the resulting principal eigenfunction $\Pfb$ in $\D$ to approximate the exact stationary distribution $\Pf$ in $\D^\infty$.
We show in Theorem \ref{theorem: artificial boundary condition} that $\Pfb$ converges toward $\Pf$ as $\D \rightarrow \D_\infty$.
We also use a numerical experiment (Fig. \ref{fig: birth-death}(b)) in Section 7.1 to show the convergence rate of the error between $\Pfb$ and $\Pf$ within $\D$ agrees well with the maximum gradient at boundaries, i.e., $\max_{\bfx \in \partial \D} | \nabla \Pfb (\bfx)|$.

\paragraph{Discretization error}

In step {\bf (a3)}, the TPA uses the finite difference approximations of the (elliptic) Dirichlet eigenvalue problem. Although many authors \cite{kuttler1970finite,gary1965computing,keller1965on} have studied the convergence of difference schemes for selfadjoint eigenvalue problems, non-selfadjoint problems in higher dimensions are less studied~\cite{carasso1969finite}, because it is generally not possible to build a monotone difference scheme using a narrow stencil~\cite{feng2013recent}.
Recently, a monotone and conservative difference scheme for elliptic operators with mixed derivatives has been proposed~\cite{samarskii2002monotone,rybak2004monotone,matus2004difference}. 
The scheme is useful to discretise the CFPEs in lower dimensions, but it is difficult to formulate in tensor formats, and thus may hardly be applied to higher dimensions.
In Section 4.1, we tailor the difference scheme such that it is capable to address the high-dimensional Fokker-Planck Dirichlet eigenvalue problem, and enjoys precise tensor decomposition (see Section 5.1). 
The difference scheme generates an irreducible $M$-matrix as the assembled matrix (Lemma \ref{lemma: A_h => M-matrix}), under appropriate conditions on the stoichiometric coefficients (Remark \ref{remark: M-matrix conditions on stoichiometrics}). 
We prove in Theorem \ref{theorem fdm convergence} that it gives a second order accurate approximation of both the principal eigenvalue and eigenfunction, which is also validated numerically (Fig. \ref{fig: birth-death}(c)).

\paragraph{Tensor rounding error}

Step {\bf (a4)} of the TPA introduces the tensor representations to the traditional finite difference discretization. 
Thanks to the new compact difference scheme (in step {\bf (a3)}), we show in Section 5.1 that the assembled matrix of the CFPE operator admits exact canonical tensor representation as a sum of canonical tensor products (CP format~\cite{hitchcock1927the}), with tensor ranks bounded by $\mathcal{O}(MN^2)$ (Proposition \ref{lemma: canonical rank of FP opeartor}).
In Section 5.2, the CFPE operator in CP format is then restructured into TT format under the same rank bound (Lemma \ref{lemma: canonical to TT}).
In Section 5.3, the TT matrix operator is further suppressed by the QTT representation, and Theorem \ref{theorem: CFPE ranks} gives the storage requirement for the CFPE operator in the QTT format to be of order $\mathcal{O}(N^5 M^2 \log_2 (n))$, where $n$ is the number of grid nodes in each dimension.
Moreover, once the assembled matrix is already represented in the QTT format, such storage estimate could be further reduced by an algorithm that truncates the tensor separation rank~\cite{oseledets2011tensorb}. 
The truncation introduces perturbations to the entries of the original QTT matrix, and Theorem \ref{theorem: tensor truncation error} links the perturbed tensor eigenvalue problem to the traditional matrix perturbation theory~\cite{deif1995rigorous}, and present a linear convergence of the tensor rounding error.
In the numerical experiment (Fig. \ref{fig: birth-death}(d) in Section 7.1), we find the effect of the tensor rounding could go beyond the linear region.

\paragraph{Algebraic error}

In step {\bf (a5)}, the TPA uses a tensor-structured inverse power method to search for a low-rank approximation of the principal eigenfunction in QTT format (see Algorithm \ref{algorithm: inverse power method}).
In order to legitimise this approach, we derive conditions (assumptions) in Section 6.1, such that there exist desirable low-rank approximations~\cite{beylkin2002numerical}.
We show in Proposition \ref{theorem: qtt for N-dimensional non-gaussian} that the existence of low-rank QTT approximation of a distribution is determined by the fact that whether the distribution could be expressed by a sum of a minimum number of Gaussian functions. 
These Gaussian functions should be away from the boundaries, and the boundary effects are not significant (Lemmas \ref{lemma: qtt for one-dimensional gaussian} and \ref{lemma: qtt for N-dimensional gaussian}).
Under such conditions, we show there exists a QTT $\varepsilon$-approximation with ranks bounded by $\mathcal{O}(\log (N/\varepsilon))$ (see Remark \ref{remark: QTT representation of multivariate non-Gaussian}).
It means that one saves the storage of order $\mathcal{O}(\log^2 (1/\varepsilon))$ by allowing a tensor approximation error $\varepsilon$.
The existence also legitimises our use of the truncated inverse iterations to search for low-rank QTT $\varepsilon$-approximations.
Further in Theorem \ref{theorem: algebiraic error convergence}, we show that, if one allows $\varepsilon$-approximations to all intermediate solutions of the inverse iterations, the whole procedure still converges to the exact principal eigenfunction, with the algebraic error of order $\mathcal{O}(\varepsilon)$ (Remark \ref{remark: inexactness of the inverse power method}). The estimate on the algebraic error is confirmed in the numerical experiment in Fig. \ref{fig: birth-death algebraic} of Section 7.1.

\section{Modelling error}

We will now investigate the convergence of the Fokker-Planck approximation \eqref{eq: CFPE} of the master equation \eqref{eq: CME} in the limit of large volumes $\Vol$, the so-called thermodynamic limit. 
Specifically, our main interest here is to derive the leading order error in the difference of the distribution, 
\begin{align*}
	\sup_{\bfn \in \mathbb{R}^N \cap \D_\infty} \left| \Pfs (\bfn) - \Pms (\bfn) \right|, 
\end{align*}
where $\Pms$ and $\Pfs$ are the solutions of \eqref{eq: CME} and \eqref{eq: CFPE}, respectively.
We will derive the leading order error by comparing the system size expansions (SSEs) of the CME and the CFPE.

\subsection{System-size expansion of the CME and the CFPE}

In this section, following the original development by Van Kampen \cite{vankampen1992stochastic}, we carry out the system size expansion of the time-dependent CME, 
\begin{align} \label{eq: CME time-dependent}
	\partial \Pm (\bfn, t) / \partial t = \Om (\bfn) \Pm (\bfn,t),
\end{align}
where $\Pm$ represents the time-dependent CME distribution.
Later, we will  extend the SSE to expand the CFPE.
The starting point of the SSE is performing the change of variables, or ansatz,
\begin{align} \label{eq: changeofvariable}
	\frac{n_i}{\Vol} = \phi_i + \sqrt{\Vol} \varepsilon_i,
	\quad
	i = 1,2,\ldots, N,
\end{align}
where the instantaneous molecular population $\bfn$ is decomposed into a deterministic part $\bfphi = \left[ \phi_1, \phi_2, \ldots, \phi_N \right]^T$ and a fluctuation part $\bfvarepsilon = \left[ \varepsilon_1, \varepsilon_2, \ldots, \varepsilon_N \right]^T$.
The distribution of the fluctuations $\bfvarepsilon$ is denoted by $\Pmsse (\bfvarepsilon, t)$, and in the large volume limit, the change of variable \eqref{eq: changeofvariable} implies $\Pmsse (\bfvarepsilon,t) = \Vol^{N/2} \Pm (\bfn,t)$.

The derivation of the CME in the new variables is performed by expanding the step operators and the propensity functions. 
Taylor expanding the step operator yields
\begin{align*}
	 \Os^{-\boldsymbol{\Vs}_{j}} - 1 = 
	 \sum_{i=1}^\infty	(-1)^i \Vol^{-i/2} a_j^{(i)}, 
\end{align*}
where 
$a_j^{(s)} = \left( \sum_{i=1}^\infty \Vs_{j,i} {\partial}/{\partial \varepsilon_i} \right) / s!$ for $1 \leq j \leq M$ and $s \geq 1$.
For the propensity functions, the Taylor expansion series can be written as
\begin{align} \label{eq: expansionalphafirst}
	\alpha_j ( \Vol \bfphi + \sqrt{\Vol} \bfvarepsilon ) 
	= \sum_{i=0}^\infty
	\left( \Vol \right) ^{-i/2}
	\sum^i_{\substack{c_1, c_2, \ldots , c_N = 0 \\ c_1 + c_2 + \cdots + c_N = i}} 	
	\frac{\varepsilon_1^{c_1} \cdots \varepsilon_N^{c_N}}{c_1! \cdots c_N!}
	\frac{\partial^i}{\partial \varepsilon_1^{c_1} \cdots \partial \varepsilon_N^{c_N}}
	\alpha_j \left( \Vol \bfphi \right),
\end{align}
and further the term $\alpha_j \left( \Vol \bfphi \right)$ can be expanded as
\begin{align} \label{eq: expansionofalpha}
	\alpha_j \left( \Vol \bfphi \right) = 
	\sum_{s=0} ^\infty	\Vol^{1-s} f_j ^{(2s)} \left( \bfphi \right),
\end{align}
and the coefficients $f_j^{(2s)}$ are given by
\begin{align} \label{eq: determratefunction}
	f_j^{(2s)} \left( \bfvarepsilon \right) = k_j 
	\sum^s_{\substack{c_1, c_2, \ldots , c_N = 0 \\ c_1 + c_2 + \cdots + c_N = s}} 
	\prod_{i=1} ^N 
	\phi_i ^{\Vs^-_{j,i} - c_i}	
	\begin{bmatrix}
		\Vs_{j,i}^- \\ c_i
	\end{bmatrix}
	\mathbbm{1}_{c_i < \Vs_{j,i}^-}, 
	\quad
	s = 1,2,3, \ldots, 
\end{align}
where $\left[ \begin{smallmatrix} \cdot \\ \cdot \end{smallmatrix} \right]$ denotes the Stirling number of the first kind.
Then, substituting \eqref{eq: expansionofalpha} into \eqref{eq: expansionalphafirst} yields the series expansion of the propensities, written as
\begin{align*}
	\alpha_j  ( \bfn ) = \Vol  
	\left[ 
		b_j^{(0,0)} 
		+ \Vol^{-1/2} b_j^{(1,0)} 
		+ \Vol^{-1} b_j^{(2,0)}
		+ \Vol^{-1} b_j^{(0,2)}
		+ \Vol^{-3/2} b_j^{(1,2)}
		+ \mathcal{O } 
		\left( \Vol ^{-2} \right)
	\right],
\end{align*}
where $b_j ^{(m,n)} = f_j^{(n)}\left( \bfphi \right) \left( \sum_{i=1}^N \varepsilon_i \partial / \partial \phi_i \right) / m!$ for $m=0,1,2, \ldots,$ and $n = 0,2,4, \ldots$.

Substituting the expansions \eqref{eq: expansionalphafirst} and \eqref{eq: expansionofalpha} into the CME \eqref{eq: CME time-dependent}, and rearranging the results in the inverse power of $\sqrt{\Vol}$, we have the CME in the new variables as
\begin{equation} \label{eq: CMESSE}
	\begin{aligned}
		\frac{\partial \Pmsse (\bfvarepsilon,t)}{\partial t}
		= \sum_{s=0}^\infty
			\Vol^{-s/2}
			\mathcal{L}_{\mathrm{m},s} 
			\Pmsse (\bfvarepsilon , t),
	\end{aligned}
\end{equation}
where, for notational simplicity, we define the operator $\mathcal{L}_{\mathrm{m},s}$ on order $\Vol^{-s/2}$ in the form
\begin{align*}
	\mathcal{L}_{\mathrm{m},s} =
	\sum_{j=1}^M 
	\left(
		\sum_{w=0}^{\lceil s/2 \rceil} 
		\sum_{v = 1} ^{s-2(w-1)}
		(-1)^v
		a_j^{(v)}
		b_j ^{(s-v-2(w-1),2w)}
	\right).
\end{align*}
In \eqref{eq: CMESSE}, the leading order $\Vol^{1/2}$ terms cancel out given the condition that $\bfphi$ satisfies the deterministic reaction rate equation.

Analogously, we can apply the same expansion procedure to the time-dependent CFPE, written as
\begin{align} \label{eq: CFPE time-dependent}
\partial \Pf (\bfx,t) / \partial t = \Of (\bfx) \Pf (\bfx,t).
\end{align} 
Let $\Pfsse (\bfvarepsilon,t)$ be the transformed version of the CFPE distribution $\Pf (\bfn,t)$ by the change of variables \eqref{eq: changeofvariable}.
Then, it can be shown that $\Pfsse$ satisfies the following SSE expanded equation of the form
\begin{equation} \label{eq: CFPESSE}
	\begin{aligned}
		\frac{\partial \Pfsse (\bfvarepsilon,t)}{\partial t}
		= \sum_{s=0}^\infty
			\Vol^{-s/2}
			\mathcal{L}_{\mathrm{f},s} 
			\Pfsse (\bfvarepsilon , t),
	\end{aligned}
\end{equation}
where 
\begin{align*}
	\mathcal{L}_{\mathrm{f},s} = 
	\sum_{j=1}^M 
	\left(
		\sum_{w=0}^{\lceil s/2 \rceil} 
		\sum_{v = 1} ^{\min (s-2(w-1),2)}
		(-1)^v
		a_j^{(v)}
		b_j ^{(s-v-2(w-1),2w)}
	\right).
\end{align*}

\subsection{Perturbative analysis of the modelling error}

Next, we focus on the stationary case and assume that trajectories of the deterministic reaction rate equation converge to a stable fixed point, $\lim_{t \rightarrow \infty} \bfphi ( t )  = \bfphi_0$. 
We write the stationary quantities by dropping their time dependence, i.e., $\Pmsse ( \bfvarepsilon )  = \lim_{t \rightarrow \infty } \Pmsse (\bfvarepsilon , t ) $ and $\Pfsse ( \bfvarepsilon )  = \lim_{t \rightarrow \infty } \Pfsse (\bfvarepsilon ,t)$.
Next, we perform the perturbative analysis of the CME \eqref{eq: CMESSE} and the CFPE \eqref{eq: CFPESSE} in new variables to derive the leading order error between $\Pmsse$ and $\Pfsse$.

We consider expressing distribution $\Pmsse$ in terms of a perturbation series in inverse powers of square root of system volume as 
\begin{align} \label{eq: perturbationseriesofcme}
	\Pmsse \left( \bfvarepsilon \right) = \sum_{j=0}^\infty \Pi_{\mathrm{m},j} \left( \bfvarepsilon  \right) \Vol^{-j/2}.
\end{align}
Substituting \eqref{eq: perturbationseriesofcme} into \eqref{eq: CMESSE} and equating the terms of order $\Vol^{-j/2}$ yields the equation for the expanded coefficients $\Pi_{\mathrm{m},j}$ as
\begin{align} \label{eq: eqn4expdcoefcme}
	- \mathcal{L}_{\mathrm{m},0}
	\Pi_{\mathrm{m},j} \left( \bfvarepsilon \right) 
	= \sum_{i=0}^{j-1} \mathcal{L}_{\mathrm{m}, j-i} \Pi_{\mathrm{m},i} \left( \bfvarepsilon  \right),
\end{align}
with normalisation conditions $\int_{\mathbb{R}^N} \Pi_{\mathrm{m},0} \pd \bfvarepsilon = 1$ and $\int_{\mathbb{R}^N} \Pi_{\mathrm{m},j} \pd \bfvarepsilon = 0$, $j=1,2,\ldots$.
Analogously, we write the series expansion of the CFPE distribution $\Pfsse$ as $\Pfsse \left( \bfvarepsilon  \right) = \sum_{j=0}^\infty \Pi_{\mathrm{f},j} \left( \bfvarepsilon  \right) \Vol^{-j/2}$, and the equations for the expanded coefficients $\Pi_{\mathrm{f},j}$ are given by
\begin{align} \label{eq: eqn4expdcoefcfpe}
	 - \mathcal{L}_{\mathrm{f},0} 
	\Pi_{\mathrm{f},j} \left( \bfvarepsilon \right) 
	= \sum_{i=0}^{j-1} \mathcal{L}_{\mathrm{f}, j-i} \Pi_{\mathrm{f},i} \left( \bfvarepsilon \right),
\end{align}
satisfying normalisation conditions  $\int_{\mathbb{R}^N} \Pi_{\mathrm{f},0} \pd \bfvarepsilon = 1$ and $\int_{\mathbb{R}^N} \Pi_{\mathrm{f},j} \pd \bfvarepsilon = 0$, $j=1,2,\ldots$.
Both equations \eqref{eq: eqn4expdcoefcme} and \eqref{eq: eqn4expdcoefcfpe} has the similar form where the coefficient are determined by the coefficients and operators on the lower orders, thus the leading order error between $\Pmsse$ and $\Pfsse$ can be found by comparing operators on the increasing orders.

\medskip
\begin{lemma} \label{lemma: errbetweendistoffluct}
	For sufficiently large volume size $\Vol$, we have 
		\begin{align} \label{eq: modelingerrlemma}
		\sup_{\bfvarepsilon \in \mathbb{R}^N} \left| \Pmsse (\bfvarepsilon ) - \Pfsse (\bfvarepsilon ) \right| \leq  C \Vol ^{-1/2},
		\end{align}
		where $C$ is a constant independent of $\Vol$.
\end{lemma}
\medskip

\begin{proof}
	Given the expansion series as in \eqref{eq: perturbationseriesofcme}, we can express the error between $\Pmsse$ and $\Pfsse$ in the same form as
	\begin{align} \label{eq: expandseriesoffluctuations}
		\Pmsse (\bfvarepsilon ) - \Pfsse (\bfvarepsilon) 
		= \sum_{j=0}^\infty 
		\left[ \Pi_{\mathrm{m},j} ( \bfvarepsilon ) - \Pi_{\mathrm{f},j} ( \bfvarepsilon ) \right]
		\Vol^{-j/2},
	\end{align}
where $\Pi_{\mathrm{m},j}$ and $\Pi_{\mathrm{f},j}$ satisfy \eqref{eq: eqn4expdcoefcme} and \eqref{eq: eqn4expdcoefcfpe}, respectively.
For $j=0$, the right-hand sides of both equations equal to zero and thus we have $\Pi_{\mathrm{m},0} = \Pi_{\mathrm{f},0}$.
For $j=1$, the difference between the operators follows
$\mathcal{L}_{\mathrm{m},1} - \mathcal{L}_{\mathrm{f},1} = - a_j ^{(3)} b_j ^{(0,0)}$, which is non-zero in general, and thus the coefficients do no equal, i.e., $\Pi_{\mathrm{m},1} \neq \Pi_{\mathrm{f},1}$.
Hence, the higher order error can be bounded by $O \left( \Vol^{-1/2} \right)$, which guarantees the convergence in \eqref{eq: modelingerrlemma}.
\end{proof}

Given the difference between the distributions of the fluctuations in \eqref{eq: modelingerrlemma} of Lemma \ref{lemma: errbetweendistoffluct}, one can back up the error between the CFPE distribution $\Pfs$ and  the CME distribution $\Pms$, using the scaling relationship in \eqref{eq: changeofvariable}.

\medskip
\begin{theorem} \label{theorem: modellingerrgeneral}
	For sufficiently large volume size $\Vol$, we have 
	\begin{align} \label{eq: modellingerrgeneral}
		\sup_{\bfn \in \mathbb{R}^N \cap \D_\infty} \left| \Pms (\bfn ) - \Pfs (\bfn ) \right| \leq  C \Vol ^{-(N+1)/2},
	\end{align}
where $N$ is the number of chemical species and $C$ is a constant independent of $\Vol$.
\end{theorem}
\medskip

\begin{proof}
	The modelling error \eqref{eq: modellingerrgeneral} can be derived by substituting the relation, $\Pmsse (\bfvarepsilon) - \Pfsse (\bfvarepsilon) = \Vol^{N/2} \left[ \Pms (\bfn ) - \Pfs (\bfn) \right]$, which is implied from the change of variables in \eqref{eq: changeofvariable}, into \eqref{eq: modelingerrlemma}.
\end{proof}

\subsection{Systems with  detailed balanced condition}

The error convergence \eqref{eq: modellingerrgeneral} of Theorem \ref{theorem: modellingerrgeneral} applies to the general types of reaction networks \eqref{eq: mass-action reactions} that fulfil the assumption that the trajectories of the deterministic reaction rate equation converge to a single fixed point. 
Next, we consider a type of networks obeying the detailed balanced condition.

\medskip
\begin{definition} \label{definition: detailedbalancedcondition}
	A chemical reaction network is called reversible, if for every reaction of the form \eqref{eq: mass-action reactions} with a positive reaction rate $k_{j+}>0$, there exists a backward reaction with a positive reaction rate $k_{j-}>0$.
	Further, a reversible reaction network is detailed balanced, if for each pair of reversible reactions, we have 
	\begin{align} \label{eq: detailbalancerre}
		f_{j+}^{(0)} \left( \bfvarepsilon \right) = f_{j-}^{(0)} \left( \bfvarepsilon \right),
		\quad
		1 \leq j \leq M/2,
\end{align}	
where $f_{j}^{(0)}$ are the deterministic rate functions defined in \eqref{eq: determratefunction} with the plus and minus signs referring to the forward and backward reactions, respectively.
\end{definition}
\medskip

Applying the condition \eqref{eq: detailbalancerre} of Definition \ref{definition: detailedbalancedcondition} to the equations \eqref{eq: eqn4expdcoefcme} and \eqref{eq: eqn4expdcoefcfpe} of the expansion coefficients, one can derive a higher order convergence between the CME and CFPE distributions.

\medskip
\begin{theorem} \label{theorem: modellingerrdetailbalance}
	Assuming the detailed balanced condition in Definition \ref{definition: detailedbalancedcondition} is satisfied, then for sufficiently large volume $\Vol$, we have
	\begin{align} \label{eq: modellingerrdetailbalance}
		\sup_{\bfn \in \mathbb{R}^N \cap \D_\infty} \left| \Pms (\bfn ) - \Pfs (\bfn ) \right| \leq  C \Vol ^{-(N+2)/2},
	\end{align}
where $C$ is a constant independent of $\Vol$.
\end{theorem}
\medskip

\begin{proof}
	We have shown in Lemma \ref{lemma: errbetweendistoffluct} that $\Pmsse$ and $\Pfsse$ agree on order $\Vol^0$, and now we prove they also agree on order $\Vol^{-1/2}$ given the detailed balanced condition.
	The difference between the operators $\mathcal{L}_{\mathrm{m},1}$ and $\mathcal{L}_{\mathrm{f},1}$ is given by
	\begin{align*}
		\mathcal{L}_{\mathrm{m},1} - \mathcal{L}_{\mathrm{f},1}  
		& = - \sum_{j=1}^M a_j^{(3)} b_j^{(0,0)}
		= \frac{1}{6} \sum_{j=1}^M f_j^{(0)} \left( \bfvarepsilon \right) \sum^3_{\substack{c_1, c_2, \ldots , c_N = 0 \\ c_1 + c_2 + \cdots + c_N = 3}} 	 \prod_{i=1}^N \nu_{j,i}^{c_i}
		\\
		& = \frac{1}{6} \sum_{j=1}^{M/2} \left[ f_{j+}^{(0)} \left( \bfvarepsilon \right) -  f_{j-}^{(0)} \left( \bfvarepsilon \right)  \right] \sum^3_{\substack{c_1, c_2, \ldots , c_N = 0 \\ c_1 + c_2 + \cdots + c_N = 3}} 	 \prod_{i=1}^N \nu_{j+,i}^{c_i},
	\end{align*}
where the last formula is derived from the detailed balanced assumption that the reactions come in pairs and the fact that the stoichiometric coefficients for the backward reaction is the same as the forward one but with the opposite sign.
Substituting the condition \eqref{eq: detailbalancerre} into the above formula yields $\mathcal{L}_{\mathrm{m},1} - \mathcal{L}_{\mathrm{f},1} = 0$, and from \eqref{eq: expandseriesoffluctuations}, we can derive 
\begin{align*}
		\sup_{\bfvarepsilon \in \mathbb{R}^N} \left| \Pmsse (\bfvarepsilon ) - \Pfsse (\bfvarepsilon ) \right| \leq  C \Vol ^{-1},
\end{align*}
which is an analogy of Lemma \ref{lemma: errbetweendistoffluct}. 
Then as before, reversing back the change of variables in \eqref{eq: changeofvariable} gives the convergence in \eqref{eq: modellingerrdetailbalance}.
\end{proof}

\section{Artificial boundary error}

For the purpose of computation, the stationary CFPE \eqref{eq: CFPE} is only considered in a bounded domain, $\D \subset \Omega_\infty$, and the stationary solution is approximated by the positive principal eigenfunction,
\begin{align} \label{eq: CFPE eigenvalue Omega}
    \Ofb ( \bfx) \Pfb (\bfx) &= \lambda_\D 	\Pfb ( \bfx ), \qquad \mathrm{for} \ \  \bfx \in \D	,
\end{align}
where $\Ofb (\bfx)$ denotes the truncated CFPE operator within $\D$, and $
\lambda_\Omega$ represents the principal eigenvalue.
We consider the homogeneous Dirichlet boundary condition as the artificial boundary condition on $\partial \D$, i.e.,
\begin{align*}
    \Pfb ( \bfx ) &= 0 , \qquad \mathrm{for} \ \  \bfx \in \partial \D. 
\end{align*}
Since $\D$ is a subset of $\D_\infty$, the ellipticity condition is satisfied in $\D$.
In fact, it can be shown that the sufficient conditions for fulfilling the ellipticity condition \eqref{eq: ellitic conditions in Omega^infty} are (a) $\alpha_{i,j}(\bfx) \geq c_1 > 0$ for all $\bfx \in \D$ and all $i,j=1,2,\dots,N$ and (b) there is $N$ (out of $M$) linearly independent rows of the stochiometric matrix $\boldsymbol{\nu}$.

We show in the following theorem that $\Pfb$ in \eqref{eq: CFPE eigenvalue Omega} converges to $\Pf$ in \eqref{eq: CFPE} as $\D \rightarrow \mathbb{R}^N$.

\medskip
\begin{theorem} \label{theorem: artificial boundary condition}
	If the stationary CFPE \eqref{eq: CFPE} admits an bounded solution $\Pfs$ in $\D_\infty$, which is unique up to a constant multiplication. Then, for $\D \subset \D_\infty$, we have 
	\begin{align*}
		\lambda_\D < 0, 
		\quad \mathrm{and} \quad
		\lambda_\D \uparrow 0 
		\ \mathrm{as} \ 
		\D \rightarrow \D_\infty.
	\end{align*}
If further both $\Pfs$ and $\Pfb$ are normalised so that $\Pfs ( \bfx ) = \Pfb ( \bfx ) = 1$ for a given $\bfx \in \D$, we have the convergence $\Pfb \rightarrow \Pfs$ in $\D$ as $\D \rightarrow \D_\infty$.
\end{theorem}
\medskip

\begin{proof}
Assuming the existence and uniqueness of solution $\Pfs$ in \eqref{eq: CFPE} implies a zero (generalised) principal eigenvalue of $\Of$ in $\Omega_\infty$.
Using Proposition 2.3(iv) in \cite{berestycki2014generalizations}, we obtain a negative principal eigenvalue $\lambda_\D < 0$ of operator $\Of$ in $\D$, and its convergence $\lambda_\D \rightarrow 0$ as $\D \rightarrow \D_\infty$.
The convergence of the eigenfunctions is given in the proof of Theorem 1.4 in \cite{berestycki2014generalizations}.
\end{proof}

\section{Discretization error}

In this section we present and analyse a monotone difference scheme for the (generalized) Fokker-Planck eigenvalue problem \eqref{eq: CFPE eigenvalue Omega}.
We follow the traditional matrix-vector-based routine in the current section where the nodal points are enumerated by one index and the corresponding discretised functions are stacked into a ``long'' vector, but we will consider storage compression by tensor representation in section 5.

\subsection{Difference scheme}

In an $N$-dimensional hypercube $\D = \mathcal{I}_1 \times \cdots \times \mathcal{I}_N$, $\mathcal{I}_d = [a_d, b_d]$, we consider the uniform grid $\omega_h$:
\begin{align*}
	\omega_h = \{ \bfx_{i_1,\ldots,i_N} = ( x_{1,i_1}, \ldots, x_{N,i_N}):  x_{d,i_d} = a_d + i_d h_d,  i_d = 1,\ldots, n_d, d = 1, \ldots, N \}
\end{align*}
with constant grid step $h_d = (b_d - a_d)/(n_d+1)$.
The set of boundary grid points is $\tilde{\omega}_h = \{ \bfx_{i_1,\ldots,i_N} = ( x_{1,i_1}, \ldots, x_{N,i_N}): i_d = 0 \ \mathrm{or}\  n_d+1, d= 1, \ldots N \}$.
Let us consider the following notation conventions of difference operator at grid $\omega_h$:
\begin{equation} \label{eq: finite difference scheme details}
\begin{aligned} 
	& w^{(\pm 1d)} = w (x_{1,i_1}, \ldots , x_{d, i_d} \pm h_d, \ldots, x_{N, i_N}),
	 \\
	& w = w ( x_{1,i_1} , \ldots , x_{N,i_N}),
	\quad
	w_{x_d} = \frac{w^{(+1d)} - w}{h_d},
	\quad
	w_{\bar{x}_d} = \frac{w - w^{(-1d)}}{h_d},
	 \\
	& \Lambda_{d_1, d_2}^+(p) = \frac{((\Cdif^+_{d_1,d_2} p)_{x_{d_1}})_{x_{d_2}} + ((\Cdif^+_{d_1,d_2} p)_{\bar{x}_{d_1}})_{\bar{x}_{d_2}}}{2},
	\quad 
	d_1 \neq d_2, \ \ d_1, d_2 = 1, \ldots, N,
	 \\
	& \Lambda_{d_1, d_2}^-(p) = \frac{((\Cdif^-_{d_1,d_2} p)_{x_{d_1}})_{\bar{x}_{d_2}} + ((\Cdif^-_{d_1,d_2} p)_{\bar{x}_{d_1}})_{x_{d_2}}}{2},
	\quad 
	d_1 \neq d_2, \ \ d_1, d_2 = 1, \ldots, N,
	 \\
	&  \Lambda_d(p) = \frac{(\Cdrif_d p)_{x_d} + (\Cdrif_d p)_{\bar{x}_d}}{2},
	\quad
	\Lambda_{d,d}(p) = ( ( \Cdif_{d,d} p)_{\bar{x}_d} )_{x_d}, 
	\quad
	d = 1, \ldots, N
\end{aligned}
\end{equation}
where $\Cdif_{d_1, d_2}^+$ and $\Cdif_{d_1, d_2}^-$ represent the partial sums of the positive and negative terms in \eqref{eq: continuous FP eigenvalue problem diffusion drift coefficients}, respectively, i.e.,
\begin{align} \label{eq: finite difference scheme alpha + - }
	\Cdif_{d_1, d_2}^+ = \frac{1}{2} \sum_{r=1}^M \Vs_{r, d_1} \Vs_{r, d_2} \Cdif_{d_1, d_2} \cdot \mathbbm{1}_{\Vs_{r, d_1} \Vs_{r, d_2} >0}, 
	\quad 
	\Cdif_{d_1, d_2}^- =  \Cdif_{d_1, d_2} - \Cdif_{d_1, d_2}^+.
\end{align}

Then, the finite difference approximation of \eqref{eq: CFPE eigenvalue Omega} of second order reads
\begin{align} \label{eq: finite difference scheme}
	A_h p_h = \sum_{d=1}^N (\Lambda_d(p_h) + \Lambda_{d,d}(p_h)) + \sum_{\substack{
	d_1, d_2 = 1 \\
	 d_1 \neq d_2
}}^N (\Lambda_{d_1, d_2}^+(p_h) + \Lambda_{d_1, d_2}^-(p_h))
	=  \lambda_h p_h,
\end{align}
where $A_h$ here denotes a huge multi-dimensional matrix, and $(\lambda_h, p_h)$ is the associate principal eigenpair of $A_h$. The stencil of difference scheme \eqref{eq: finite difference scheme details} is compact, that uses only $(3^N-1)$ surrounding nodes for its discretization in $N$ dimensions. 
For investigation of a priori error estimate of the solution $p_h$, the following result has been proved in Theorem 3.7 of  \cite{liao2016high}, and we presented in the following lemma.

\medskip
\begin{lemma} \label{lemma: A_h => M-matrix}
Let us suppose that, for sufficiently small grid size $h_d$, $d=1, \ldots, N$, the following conditions are satisfied:
\begin{align} \label{eq: condition A_h M-matrix h}
	\frac{a_{d,d}}{h_d}
	 \geq 
	 \sum_{\substack{d'=1 \\ d' \neq d}}^N 	\frac{ a_{d, d'}^+ + a_{d', d}^+ - ( a_{d, d'}^- + a_{d', d}^-)}{ 2 h_{d'}} 
	 \quad
	 \mathrm{for} \ \mathrm{all} \ 
	 \bfx \in \omega_h,
\end{align}
then the stiffness matrix $(- A_h)$ in  \eqref{eq: finite difference scheme} is an irreducible $M$-matrix, and the principal eigenpair $(\lambda_h, p_h)$ of $A_h$ has the following properties: (i) $\lambda_h$ is real and algebraically (and geometrically) simple; (ii) $Re(\lambda_h^\ell) < 0$ and $| \lambda_h^\ell | \geq |\lambda_h | $ for every eigenvalue $\lambda_h^\ell$; and (iii) $p_h(\bfx) \geq 0$ for all $\bfx \in \omega_h$.
\end{lemma}
\medskip

\begin{remark} \label{remark: M-matrix conditions on stoichiometrics}
{\rm
Lemma \ref{lemma: A_h => M-matrix} suggests that the difference scheme \eqref{eq: finite difference scheme} satisfies the discrete maximum principle, and is stable and convergent. Note that, if stoichiometric coefficients of \eqref{eq: mass-action reactions} satisfy the following inequality
\begin{align}
	\sum_{j=1}^M 2\Vs^2_{j,i} \geq \sum_{\substack{i'=1 \\ i' \neq i}}^N \sum_{j=1}^M |\Vs_{j,i} \Vs_{j,i'}|, \quad
	\mathrm{for }\ 
	i = 1, 2, \ldots, N,
\end{align}
then one can choose $h_1 = h_2 = \ldots = h_N = h$ which guarantees that condition \eqref{eq: condition A_h M-matrix h} is satisfied. To fully satisfy the condition of Lemma \ref{lemma: A_h => M-matrix}, we further require $h$ small enough such that the drift terms $\Lambda_d$ in \eqref{eq: finite difference scheme} are small enough to keep the positivity of the off-diagonal entries in $A_h$.
}
\end{remark}

\subsection{Convergence of discretization error}

Let $0> \lambda_h \geq \lambda^2_h \geq \lambda^3_h \geq \cdots$ be the eigenvalues of $A_h$, considered as approximations to the first $(n_1\times \cdots \times  n_N)$ eigenvalues of the differential operator $\Ofb$ in \eqref{eq: CFPE eigenvalue Omega}. For a positive integer $1 \leq j \leq n_1\cdots n_N$ we define $p_h^j$ to be the eigenvector corresponding to $\lambda^j_h$, normalised so that $\left\| p_h^j \right\|_1/\prod_i \gridh_i = 1$. We will show the convergence rate of $\lambda_h^1 $ and $p_h $ to the continuous problem in \eqref{eq: CFPE eigenvalue Omega} as the grid size $h_i \rightarrow 0$, $i= 1,\ldots, N$, in the following theorem.

\medskip
\begin{theorem} \label{theorem fdm convergence}
	Let $\lambda_h$, $p_h$ be principal eigenpairs of $A_h$ with $\left\| p_h \right\|_1=1$. 
	Let $\Pfb$ be an eigenfunction of $\Ofb$ associated with $\lambda_\Omega$, and let $p_\Omega$ be the vector obtained from $\Pfb$ by grid point evaluation in $\omega_h$. Assume $p_\Omega$ normalised so that $p_h^T p_\Omega = 1$. 
	Define $h \geq h_d$ for $d=1, \ldots, N$, that satisfies the conditions of Lemma \ref{lemma: A_h => M-matrix} then as $h \rightarrow 0$, we have
	\begin{align} \label{eq: theorem fdm eigenvalue convergence}
		| \lambda_\Omega - \lambda_h | \leq C_1 h^2
		\quad
		\mathrm{and}
		\quad
		\| p_\Omega - p_h \|_\infty  \leq C_2 h^2, 
	\end{align}
where $C_1$, $C_2$ are positive constants.
\end{theorem}
\medskip

\begin{proof}
	Because the difference scheme \eqref{eq: finite difference scheme} is properly centered and we assume sufficient smoothness of $\Pfb(\bfx)$, we have
\begin{align} \label{eq: proof theorem: fdm convergence 1}
	A_h p_\Omega + \tau_h =  \lambda_\Omega p_\Omega,
\end{align}
where $\tau_h$ is the truncation error of the difference scheme. One can show, by using Taylor expansion in space~\cite{samarskii2002monotone}, that 
\begin{align} \label{eq: proof theorem: fdm convergence 3}
	 \| \tau_h \|_\infty \leq \bar{C}_0 h^2,
\end{align}
where $\bar{C}_0$ is a constant depending on $p_\D$.
Now we write $p_\Omega$ as a linear combination of the eigenvectors of $A_h$, i.e., 
\begin{align} \label{eq: proof theorem: fdm convergence 4}
p_\Omega = \sum_{j} c_j p_h^j.
\end{align}
Substituting into \eqref{eq: proof theorem: fdm convergence 1} gives
\begin{align} \label{eq: proof theorem: fdm convergence 2}
	\tau_h = ( \lambda_\Omega - A_h) p_\Omega = \sum_j c_j(\lambda_\Omega - \lambda_h^j) p_h^j.
\end{align}
Let $q_h$ be the normalised principal left eigenvector of $A_h$ corresponding to $\lambda_h$, then multiply $q_h^T$ to both sides of the equation \eqref{eq: proof theorem: fdm convergence 2} and use the relation in \eqref{eq: proof theorem: fdm convergence 3}, we have 
\begin{align} \label{eq: proof theorem: fdm convergence 5}
	\left| c_1 (\lambda_\Omega - \lambda_h) \right| 
	\leq 
	\left| q_h^T \tau_h \right| 
	\leq \| q_h \|_1 \|\tau_h\|_\infty \leq \bar{C}_1 h^2,
\end{align}
where $\bar{C}_1$ is a constant depending only on $\Cdif_{i,j}$, $\Cdrif_i$, $i,j=1, \ldots N$, and $p_\D$. 
We may assume $c_1 \neq 0$, and divide both sides of the inequality by $|c_1|$. This gives the convergence rate of the principal eigenvalue in \eqref{eq: theorem fdm eigenvalue convergence}.

Now we derive the convergence of the eigenvector $p_h$ towards $p_\Omega$. Let $Q$ be a matrix formed by the columns of the right eigenvectors of $A_h$, then the matrix form of \eqref{eq: proof theorem: fdm convergence 4} reads $p_\Omega = Q c$, where $c$ denotes a column vector with entries $c_j$. Similarly, the matrix form of \eqref{eq: proof theorem: fdm convergence 2} is given by 
\begin{align}\label{eq:tauhexpression}
\tau_h = (\lambda_\Omega - A_h) Qc = Q \Phi_h  c,
\end{align}
where $\Phi_h$ is a diagonal matrix with $( \lambda_\Omega - \lambda_h^j )$, $j=1, \ldots, n$, as the diagonal entries.
From Lemma \ref{lemma: A_h => M-matrix}, $\lambda_h$ is a simple eigenvalue of $A_h$, and $p_h$ is linearly independent with $p_h^j$, $j=2, 3, \ldots$. 
Thus we can define a transformation that makes  $p_h^j$, $j=2, 3, \ldots$, orthogonal to $p_h$.
We define matrix $Z = [z_{i,j}]_{n\times n}$ as
\begin{align*}
Z = 
\begin{pmatrix}
  1 & -\frac{\langle p_h, p_h^2 \rangle}{\langle p_h, p_h \rangle} & \cdots & -\frac{\langle p_h, p_h^n \rangle}{\langle p_h, p_h \rangle} \\
  0 & 1 & 0 & 0 \\
  0 & 0 & \ddots & 0 \\
  0 & \cdots & 0 & 1
 \end{pmatrix}
 _{n \times n},
\end{align*}
where $\langle p_h, p_h^j \rangle$ denotes the inner product of vectors $p_h$ and $p_h^j$. Let $\tilde{Q} = QZ$, then the first column of $\tilde{Q}$ is orthogonal to all other columns.
Then, multiplying $\tilde{Q}^T$ to both sides of \eqref{eq:tauhexpression} gives
\begin{align} \label{eq: proof theorem: fdm convergence 6}
	\tilde{Q}^T \tau_h = \tilde{Q}^T \tilde{Q} Z^{-1}   \Phi_h c.
\end{align}
From the definition of $\tilde{Q}$, the structure of $\tilde{Q}^T \tilde{Q}$ is of the form
\begin{align*}
\tilde{Q}^T \tilde{Q} = 
	\begin{pmatrix}
	1 & \mathbf{0}^T \\
	\mathbf{0} & \bar{Q}^T \bar{Q}
	\end{pmatrix}
	_{n \times n},
\end{align*}
where $\bar{Q}$ is an submatrix of $\tilde{Q}$ with the first column vector removed. 
It can also be verified that
\begin{align*}
	\tilde{Q}^T \tilde{Q} Z ^{-1} = 
	\begin{pmatrix}
		1 & - z_{1, 2:n} \\
		\mathbf{0} & \bar{Q}^T \bar{Q}
	\end{pmatrix}
	_{n \times n},
\end{align*}
where $z_{1,2:n}$ refers to a vector formed by the first row of matrix $Z$ except the first element. 
Further, since $ \Phi_h$ is diagonal matrix, we can reduce \eqref{eq: proof theorem: fdm convergence 6} to 
\begin{align*}
	\bar{Q}^T \tau_h = \bar{Q}^T \bar{Q} \bar{\Phi}_h \bar{c},
\end{align*}
where $\bar{\Phi}_h$ is a submatrix of $\Phi_h$ with the first row and the first column removed, and $\bar{c}$ is a sub-vector of $c$ with entries $(c_2, \ldots, c_n) ^T$.
Then, we have
\begin{align} \label{eq: proof theorem: fdm convergence 7}
	\left\| \bar{c} \right\|_\infty 
	\leq \left\| \bar{\Phi}_h^{-1} \right\|_\infty \left\| (\bar{Q}^T \bar{Q})^{-1} \bar{Q}^T \right\|_\infty \| \tau_h \|_\infty
	\leq \bar{C}_2 \left\| \bar{\Phi}_h^{-1} \right\|_\infty h^2,
\end{align}
with constant $\bar{C}_2 > 0$, where we have assumed that eigenvectors $p_h^j$ have bounded entries, i.e., $\left\| p_h^j \right\|_\infty < \infty$. In above inequality, we notice that $\left\| \bar{\Phi}_h^{-1} \right\|_\infty = \max_{j=2,\ldots,n} 1 / | \lambda_\Omega - \lambda_h^j|$.
Since we have proved the convergence of the eigenvalue $\lambda_h$ towards $\lambda_\Omega$, for sufficiently small $h$, there exists a positive constant $\tilde{C}_2$, such that
\begin{align*}
	\inf_{j \neq 1} \left| \lambda_\Omega - \lambda_h^j \right| \geq \tilde{C}_2 >0.
\end{align*}
We can further bound $\| \bar{c} \|_\infty$ in \eqref{eq: proof theorem: fdm convergence 7} by
\begin{align*}
	\| \bar{c} \|_\infty \leq \frac{\bar{C}_2} {\tilde{C}_2} h^2 = C'_2 h^2,
\end{align*}
where $C'_2$ is a constant.
We may need to assume the eigenvectors, $p_h^j$, $j=1, 2, \ldots, n$, are bounded. It then follows that 
\begin{align*}
	\left\| p_\Omega - p_h \right\|_\infty 
	= \left\| \sum_{j=1}^n c_j p_h^j - p_h \right\|_\infty
	\leq \max_{j=2, \ldots, n} | c_j | \left\| p_h^j \right\|_\infty
	\leq C_2 h^2,
\end{align*}
where $C_2$ is a constant. The above inequality gives the convergence of principal eigenvector in \eqref{eq: theorem fdm eigenvalue convergence}.
\end{proof}

\section{Tensor-structured approximation}

In this section we discuss the tensor representation of the difference scheme \eqref{eq: finite difference scheme}, and the associate error caused by the separation rank truncations.

\subsection{Canonical tensor products applied to Fokker-Planck problem}

The \emph{tensor product} (Kronecker product, direct product) of two matrices $A = \begin{pmatrix} a_{i,j} \end{pmatrix} _{n\times n}$ and $B = \begin{pmatrix} b_{i,j} \end{pmatrix}_{m \times m}$, denoted by $A \otimes B$, can be written as a matrix in block partition form
\begin{align}
	A \otimes B = 
	\begin{pmatrix}
	a_{1,1} B & \cdots & a_{1,n} B \\
	\vdots & & \vdots \\
	a_{n,1} B & \cdots & a_{n,n} B
	\end{pmatrix}_{nm \times nm}
\end{align}
A detailed account of properties of tensor product is given in \cite{halmos1974finite}. Some of the elementary properties are:
\begin{align*}
	(A + B ) \otimes C = A \otimes C + B \otimes C, \quad (A \otimes B) (C \otimes D) = AC \otimes BD.
\end{align*}
For brevity, we do not indicate explicitly the sizes of the matrices involved; we assume throughout that the sizes of matrices and vectors are compatible with the indicated operations. We will refer the tensor product of matrices as rank-1 tensor matrix, and sum of rank-1 tensor matrices as rank-$n$ matrix, both denoted in bold font capitals.

For compact finite difference scheme \eqref{eq: finite difference scheme details}, it is readily verified that the upwind difference and the downwind difference are, respectively, of the form
\begin{align} \label{eq: tensor product for difference operator}
	\mathbf{D}_{d} = I \otimes \cdots \otimes D_{d} \otimes \cdots \otimes I, 
	\ \ \mathrm{and} \ \ 
	\mathbf{D}_{\hat{d}} = I \otimes \cdots \otimes D_{\hat{d}} \otimes \cdots I
\end{align}
for $d, \hat{d} = 1,2,\dots,\Ns$,
where $I$ denotes identity matrix of appropriate sizes.
The upwind difference matrix $D_d$ has entries $1/h$ and $-1/h$ distributed along its super-diagonal and diagonal, and the downwind difference matrix $D_{\hat{d}}$ has them alone its diagonal and sub-diagonal.
It follows that the tensor representation of the difference operator in \eqref{eq: finite difference scheme details} reads
\begin{align} \label{eq: tensor product of the sub operators}
	\mathbf{\Lambda}_d & = \frac{1}{2} \mathbf{D}_d \mathbf{F}_d + \frac{1}{2} \mathbf{D}_{\hat{d}} \mathbf{F}_d,
	\quad 
	&\mathbf{\Lambda}^+_{d_1, d_2} &= \frac{1}{2} \mathbf{D}_{d_1} \mathbf{D}_{d_2} \mathbf{G}^+_{d_1,d_2} + \frac{1}{2} \mathbf{D}_{\hat{d}_1} \mathbf{D}_{\hat{d}_2} \mathbf{G}^+_{d_1,d_2}, 
	\nonumber \\
	\mathbf{\Lambda}_{d,d} & = \mathbf{D}_d \mathbf{D}_{\hat{d}} \mathbf{G}_{d,d}, 
	\quad \mathrm{and} \quad	
	&\mathbf{\Lambda}^-_{d_1, d_2} &= \frac{1}{2} \mathbf{D}_{d_1} \mathbf{D}_{\hat{d}_2} \mathbf{G}^-_{d_1,d_2} + \frac{1}{2} \mathbf{D}_{d_1} \mathbf{D}_{\hat{d}_2} \mathbf{G}^-_{d_1,d_2},
\end{align}
for $d, d_1,d_2, \hat{d}_1, \hat{d}_2 = 1,2,\dots,\Ns$,
where $\mathbf{G}^+_{d_1, d_2}$ and $\mathbf{G}^-_{d_1, d_2}$ are the tensor matrix representations of the positive and negative summation of diffusion coefficients, $a^+_{d_1,d_2}$ and $a^-_{d_1,d_2}$, in \eqref{eq: finite difference scheme alpha + - } of the form
\begin{align*}
	\mathbf{G}^+_{d_1,d_2} = \sum_{r=1}^M \Vs_{r,d_1} \Vs_{r,d_2}  \mathbbm{1}_{\Vs_{r,d_1} \Vs_{r,d_2}>0} \mathbf{H}_r, 
	\quad
	\mathbf{G}^-_{d_1,d_2} = \sum_{r=1}^M \Vs_{r,d_1} \Vs_{r,d_2}  \mathbbm{1}_{\Vs_{r,d_1} \Vs_{r,d_2}<0} \mathbf{H}_r, 
\end{align*}
and $\mathbf{F}_d$ is the tensor representation of the drift coefficients, $b_d$, in \eqref{eq: continuous FP eigenvalue problem diffusion drift coefficients} of the form
\begin{align*}
	\mathbf{F}_{d} = \sum_{r=1}^M \Vs_{r,d} \mathbf{H}_r, 
\end{align*}
where $\mathbf{H}_r$ is a rank-1 representation of propensities, $\alp_r$, by
\begin{align} \label{eq: propensity in tensor product}
	\mathbf{H}_r = H^r_1 \otimes \cdots \otimes H^r_N,
\end{align}
for $r=1,2,\ldots, M$, and
\begin{align*}
	H_j^r (\ell_1^j, \ell_2^j) = \left\{ 
  \begin{array}{l l}
    (\Vs_{r,j}^-)! \beta_{j,r}(x_{j,\ell_1^j})  
         & \quad \ell_1^j = \ell_2^j, \\
    1 & \quad \ell_1^j \neq \ell_2^j,
  \end{array} \right.
\end{align*}
for  $\ell_1^j,\ell_2^j = 1,2,\ldots,n_j$, $j=1,2,\ldots, N$ and $r = 1,2,\ldots, M$,
where $\beta_{j,d}(x_{j,\ell})$ was defined in Section~1.
Then the tensor analogy of finite difference discretization of Fokker-Planck equation in \eqref{eq: finite difference scheme} reads
\begin{align} \label{eq: canonical tensor for A_h}
	\mathbf{A}_h \bfp_h \equiv  \left( \sum_{d=1}^N (\mathbf{\Lambda}_d + \mathbf{\Lambda}_{d,d}) + \sum_{\substack{
	d_1, d_2 = 1 \\
	 d_1 \neq d_2
}}^N (\mathbf{\Lambda}_{d_1, d_2}^+ + \mathbf{\Lambda}_{d_1, d_2}^-) \right) \bfp_h 
	= - \lambda_h \bfp_h,
\end{align}
where $\mathbf{A}_h$ is an exact permuted reformulation of $A_h$ in \eqref{eq: finite difference scheme} as a sum of rank-1 tensor matrices, and $\bfp_h$ stands for the canonical tensor representation of the ``long'' vector $p_h$.
The separation rank bound of $\mathbf{A}_h$ is given as follows..

\begin{proposition} \label{lemma: canonical rank of FP opeartor}
	For the compact difference scheme \eqref{eq: finite difference scheme} for an $N$-dimensional Fokker-Planck equation \eqref{eq: CFPE eigenvalue Omega}, 
	the assembled matrix $A_h$ can be exactly expressed as rank-$R$ canonical tensor
	with the rank bounded by $R \leq 2 MN+2N^2$, where $M$ and $N$ are the numbers of reactions and species, respectively, defined in \eqref{eq: mass-action reactions}.
\end{proposition}

The rank bound can be directly derived from the explicit structure of $\mathbf{A}_h$ in \eqref{eq: canonical tensor for A_h}.
It scales quadratically in $N$ and linearly in $M$, that indicates the storage requirement of the assembled canonical tensor matrix scales as $\mathcal{O}(n^2MN^2)$, for $n \geq n_d$, $d=1, \ldots , N$.

\subsection{Tensor train representation}

The tensor train (TT) representation of the canonical matrix $\mathbf{A}_h$ can be described as
\begin{align} \label{eq: tt matrix}
 	\mathbf{A}_h = \mathbf{U}^{(1)} \times_3 \mathbf{U}^{(2)} \times_{5} \cdots \times_{2N-1} \mathbf{U}^{(N)},
\end{align}
where the core tensors are defined as  
$\underline{\mathbf{U}}^{(d)} \in \mathbb{R}^{n_d \times n_d \times R_{d}}$ 
for $d = 1$,  
$\underline{\mathbf{U}}^{(d)} \in \mathbb{R}^{R_{d-1} \times n_d \times n_d}$ 
for $d = N$, and 
$\underline{\mathbf{U}}^{(d)} \in \mathbb{R}^{R_{d-1} \times n_d \times n_d \times R_{d}}$ for $d = 2, \ldots, N-1$. 
The generalised mode-$d$ product of  two TT matrices,
$\mathbf{A} \in \mathbb{R}^{n_1 \times \cdots \times n_d}$ 
and 
$\mathbf{B} \in \mathbb{R}^{n_d \times \cdots \times n_{N}}$,
yields TT matrix 
$\mathbf{C} = \mathbf{A} \times_d \mathbf{B} \in \mathbb{R}^{n_1 \times \cdots n_{d-1} \times n_{d+1} \times \cdots \times n_{N}}$ 
with entries 
$\mathbf{c}_{i_1, \ldots, i_{d-1}, i_{d+1}, \ldots, i_N} = \sum_{i_d=1}^{n_d} \mathbf{a}_{i_1, \ldots, i_d} \cdot \mathbf{b}_{i_d, \ldots, i_{N}}$. 
The sizes of bridging dimensions $R_1, \ldots, R_N$ are called the \emph{ranks} of the TT matrix.
The canonical representation \eqref{eq: canonical tensor for A_h} can be directly converted to
the tensor train representation \eqref{eq: tt matrix} 
and the ranks of the resulting TT matrix are given in the following lemma, where $\|\cdot\|_F$ stands for the Frobenius norm.

\medskip
\begin{lemma}[Corollary 2.3 in \cite{oseledets2011tensor}] \label{lemma: canonical to TT}
	If a tensor matrix $\mathbf{A}$ admits a canonical representation $\mathbf{A}_c$ with rank $R$ and accuracy $\| \mathbf{A} - \mathbf{A}_c\|_F \leq \varepsilon$, then there exists a TT representation $\mathbf{A}_t$ with TT-ranks $R_d \leq R$ and accuracy $\| \mathbf{A} - \mathbf{A}_t \|_F \leq\sqrt{N-1} \varepsilon$.
\end{lemma}
\medskip

Since \eqref{eq: canonical tensor for A_h} is exact, i.e. $\varepsilon = 0$, the converted TT representation \eqref{eq: tt matrix} is exact with storage estimate as $\mathcal{O}(n^2M^2N^5)$.
Although the storage estimate appears to be smaller for the canonical representation, TT representation is more stable because it allows SVD-based algorithms~\cite{oseledets2009breaking}.

\subsection{Quantized tensor train representation} 

We have discussed using the tensor representation to break the \emph{curse of dimensionality} in $\Ns$ physical dimensions of the state space $\D$.
Now, each of the physical dimension is further quantized into several virtual dimensions, and consequently, each of the core tensors in \eqref{eq: tt matrix} is further decomposed as the product of quantized core tensors with smaller mode size.

Consider tensor $\mathbf{A}_h$ in \eqref{eq: tt matrix} with core tensors 
$\mathbf{U}^{(d)} \in \mathbb{R}^{R_{d-1} \times n_d \times n_d \times R_d} $ 
and 
$n_d = 2^{l_d}$, $l_d \in \mathbb{N}$, $d = 1, \ldots, N$, 
the mode index 
$1 \leq i_d \leq n_d$ 
can be mapped to binary representation with the quantized indices 
$i_{d, \ell} \in \{ 1, 2 \}$, $\ell = 1, \ldots, l_d$, i.e., $i_d = \sum_{\ell = 1}^{l_d}  (i_{d,\ell}-1) 2^\ell$. 
Then the quantized decomposition of core tensors 
$\underline{\mathbf{U}}^{(d)}$, $d = 2, \ldots, N-1$ 
is given by 
\begin{align}  \label{eq: qtt}
\mathbf{U}^{(d)} = \mathbf{U}_1^{(d)} \times_3 \mathbf{U}_2^{(d)} \times_4 \cdots \times_{l_d+1} \mathbf{U}_{l_d}^{(d)},
\end{align}
where 
$\mathbf{U}_\ell^{(d)} \in \mathbb{R}^{r_{d} \times 2 \times 2 \times R_{d, \ell}}$ for $\ell = 1$, 
$\mathbf{U}_\ell^{(d)} \in \mathbb{R}^{R_{d,\ell -1} \times 2 \times 2 \times R_{d, \ell}}$ for $\ell = 2, \ldots, l_d-1$, 
and 
$\mathbf{U}_\ell^{(d)} \in \mathbb{R}^{R_{d,\ell -1} \times 2 \times 2 \times R_{d+1}}$ for $\ell = l_d$. 
In case where $d = 1$ or $N$, the decompositions of the core tensors are identical to \eqref{eq: qtt}, except that 
$\mathbf{U}_1^{(d)} \in \mathbb{R}^{2 \times 2 \times R_{d, 1}}$ 
for $d=1$, and 
$\mathbf{U}_{l_d}^{(d)} \in \mathbb{R}^{R_{d, l_d -1} \times 2 \times 2}$ 
for $d=N$.

Substitute the quantized core tensors of the form \eqref{eq: qtt} into the TT representation \eqref{eq: tt matrix}, the resulting decomposition is the so-called quantized tensor train (QTT)~\cite{kazeev2012low}. An upper bound on the ranks of QTT representation for the Fokker-Planck operator, $\mathbf{A}_h$, is given in the following theorem.

\medskip
\begin{theorem} \label{theorem: CFPE ranks}
	Consider the Fokker-Planck equation \eqref{eq: CFPE eigenvalue Omega} discretised by the difference scheme \eqref{eq: finite difference scheme}. Assume the propensity function $\alp_j (\bfx)$ admits QTT decomposition of ranks bounded by $R_{\alp_j}$, then the assembled tensor matrix admits exact QTT  representation with ranks bounded by $R_d \leq {2}MN(N+1)$ for $d=1,\ldots, N$, and 
\begin{align} \label{eq: QTT rank bounds for CFPE opeartor}
R_{d,\ell} \leq 
	\sum_{j=1}^M \left( \nu_{j,d}^- +1 \right)
	\left[ 
	\sum_{i=1}^N \left( 2 \mathbbm{1}_{i \neq d} + 6 \mathbbm{1}_{i=d} \right) 
	+ 
	\sum_{\substack{ i, i' = 1 \\ i \neq i'}}^N 
	\left( 
	2 \mathbbm{1}_{i_1 \neq d \cap i_2 \neq d}
	+ 6 \mathbbm{1}_{i_1 \neq d \cup i_2 \neq d}
	\right)
	\right],
\end{align}
where $\ell = 1, \ldots , l_d - 1$, and $d=1, \ldots, N$.
\end{theorem}
\medskip

\begin{proof}
	A detailed proof of the above theorem is given in section 3.2.2.5 in \cite{liao2016high}.
\end{proof}

A slightly crude upper rank bound can be derived from Theorem \ref{theorem: CFPE ranks}. Let $\nu_{\max} = \max_{i,j} \nu_{j,i}^-$, we have 
\begin{align*}
	R_{d,\ell} \leq (\nu_{\max} + 1 ) M ( 2N^2 + 8N - 4),
\end{align*}
which suggests the QTT-rank is of order $\mathcal{O} \left( MN^2 \right)$. 
Subsequently, the QTT representation of Fokker-Planck operator has complexity estimate to be $O \left( M^2N^5  \log_2 (n) \right)$, logarithmic scaling in volume size. 

\subsection{Tensor rounding error}

Once the assembled tensor matrix $\mathbf{A}_h$ is already in the QTT representation as in \eqref{eq: qtt}, we want to have an approximation, $\mathbf{A}_t$, with the ``optimal'' ranks such that
\begin{align*}
	\| \mathbf{A}_h - \mathbf{A}_t \| \leq \varepsilon \| \mathbf{A}_h \|,
\end{align*}
where $\varepsilon$ is the required accuracy level. 
Here, $\| \bfp \|$ denotes any vector norm for vector $\bfp$, and $\| \mathbf{A} \|$ denotes the corresponding matrix norm.
Let $\hat{R}_d, \hat{R}_{d,\ell}$, $\ell = 1, \ldots, l_d$, $d =1, \ldots, N$ be the QTT-ranks of $\mathbf{A}_t$, and let $\hat{R} \geq \max\{ \hat{R}_d, \hat{R}_{d,\ell} \}$, under certain assumptions~\cite{hackbusch2005hierarchical}, the suboptimal rank bound $\hat{R}$ scales with $\varepsilon$ as
\begin{align*}
	\hat{R} = \mathcal{O} (\log^2 \varepsilon^{-1}).
\end{align*}
Such a procedure is usually called rounding (truncation or recompression), and as a consequence, the truncated ranks $\hat{R}$ may be significant lower than the rank bound given in Theorem \ref{theorem: CFPE ranks}.

The approximated elliptic eigenvalue problem after tensor rounding reads
\begin{align} \label{eq: CFPE eigenvalue problem after tensor truncation}
	\mathbf{A}_t \bfp_t \equiv (\mathbf{A}_h + \delta \mathbf{A}_h) \bfp_t = -\lambda_t \bfp_t ,
\end{align}
where $(\lambda_t, \bfp_t)$ stands for the principal eigenpair of $\mathbf{A}_t$, and $\delta \mathbf{A}_h = \mathbf{A}_t - \mathbf{A}_h$ represents the perturbation caused by tensor rounding. It is of interest here to obtain bounds for the differences $| \lambda_t - \lambda_h |$ and $\| \bfp_t - \bfp_h \| $. 
Such error bounds can be analogically obtained as the perturbation bounds for the principal eigenvalues and eigenvectors of corresponding matrices~\cite{deif1995rigorous}. We state it as following theorem.

\medskip
\begin{theorem} \label{theorem: tensor truncation error}
	Let $(\lambda_t, \bfp_t)$ and $(\lambda_h, \bfp_h)$ be the principal eigenpairs for tensor eigenvalue problems \eqref{eq: CFPE eigenvalue problem after tensor truncation} and \eqref{eq: canonical tensor for A_h}, respectively.
	If $\| \delta \mathbf{A}_h \| \leq \varepsilon \| \mathbf{A}_h \|$, for sufficiently small $\varepsilon$, then we have
\begin{align}
	| \lambda_t - \lambda_h | \leq C_1 \varepsilon, 
	\quad \mathrm{and} \quad
	\| \bfp_t - \bfp_h \| \leq C_2 \varepsilon,
\end{align}
where $C_2$, $C_2$ are positive constants.
\end{theorem}

\section{Algorithm}

In this section, we discuss solutions of the tensorised eigenvalue problem. 
First of all, in section 6.1, we show there exist low-rank tensor approximations to the principal eigenfunctions under certain assumptions.
Then, in section 6.2, we present an inverse scheme in tensor formats that aims to search for such low-rank approximations.
Finally in section 6.3,  we study the algebraic error of the proposed algorithm.

\subsection{Existence}

As an analogy of \eqref{eq: tt matrix} with \eqref{eq: qtt}, the QTT approximation of $\bfp_t \in \mathbb{R}^{n_1 \times \cdots \times n_N}$ with $n_d = 2^{l_d}$, $d=1, \ldots, N$, reads
\begin{align} \label{eq: TT for vectors}
	\bfp_t \approx \hat{\bfp}_t
	= \mathbf{u}^{(1)} \times_2 \mathbf{u}^{(2)} \times_3 \cdots \times_{N} \mathbf{u}^{(N)},
\end{align}
where $\times_d$ stands for the mode-$d$ product of two tensors, and the quantized core tensors $\mathbf{u}^{(d)}$ are of the form
\begin{align} \label{eq: QTT for vectors}
	\mathbf{u}^{(d)} = 
	\mathbf{u}^{(d)}_1 \times_2 \mathbf{u}^{(d)}_2 \times_3 \cdots \times_{l_d} \mathbf{u}^{(d)}_{l_d},
\end{align}
where 
$\mathbf{u}_\ell^{(d)} \in \mathbb{R}^{R_{d,\ell -1}  \times 2 \times R_{d, \ell}}$ 
for 
$\ell = 2, \ldots, l_d-1$, $\mathbf{u}_\ell^{(d)} \in \mathbb{R}^{r_{d} \times 2 \times R_{d, \ell}}$ 
for 
$\ell = 1$, 
and 
$\mathbf{u}_\ell^{(d)} \in \mathbb{R}^{R_{d,\ell -1} \times 2 \times R_{d+1}}$ 
for 
$\ell = l_d$. 
In case where $d = 1$ or $N$, the decompositions of the core tensors are identical to \eqref{eq: QTT for vectors}, except that 
$\mathbf{u}_1^{(d)} \in \mathbb{R}^{2 \times R_{d, 1}}$ 
for $d=1$, and 
$\mathbf{U}_{l_d}^{(d)} \in \mathbb{R}^{R_{d, l_d -1} \times 2}$ 
for $d=N$.
Before presenting any algorithm to seek such an approximation $\hat{\bfp}_t$, we are interested in understanding whether there exists a low-rank $\epsilon$-approximation, i.e., satisfying $\| \bfp_t - \hat{\bfp_t} \| \leq \epsilon$.

\subsubsection{Gaussian distributions}

To answer such a question in a general scenario, we start with the cases where the solution is a Gaussian distribution. In single-dimensional cases, the rank bounds are subject to the following lemma.

\medskip
\begin{lemma}[Lemma 2.4 \cite{dolgov2012fast}] \label{lemma: qtt for one-dimensional gaussian}
	Suppose uniform grid points $-a = x_0 < x_1 < \cdots < x_n = a$, $x_i = -a + hi$, $n = 2^l$, are given on an interval $[-a,a]$ and the vector $p_t$ is defined by its elements $p_t(i) = \exp (- x_i^2 / 2 \sigma^2), i = 0, \ldots, N$.  Suppose in addition that $ \exp (- a^2 / 2 \sigma^2) \leq \epsilon $. Then for all sufficiently small $\epsilon > 0$ there exists the QTT approximation $\hat{p}_t$ with ranks bounded as
	\begin{align*}
		R \leq C \frac{a}{\sigma} \sqrt{\log \left( \frac{1}{\epsilon} \frac{\sigma}{1 + a} \right)},
	\end{align*}
and the accuracy
\begin{align*}
	\| p_t - \hat{p}_t \| \leq \left( \frac{C}{\sigma} \sqrt{\log \left( \frac{1}{\epsilon} \frac{\sigma}{1+ a} \right)} + 1 \right) \epsilon,
\end{align*}
where $C$ is a constant does not depend on $a$, $\sigma$, $\epsilon$, or $n$.
\end{lemma}
\medskip

Since the multidimensional Gaussian function is a product of one-dimensional counterparts, its canonical separation ranks are equal to 1. 
By the triangular inequality, an error bound of the QTT approximation in $N$ dimensions can be derived. Thus, we extend Lemma \ref{lemma: qtt for one-dimensional gaussian} to the multidimensional cases in the following lemma.

\medskip
\begin{lemma}  \label{lemma: qtt for N-dimensional gaussian}
Consider an $N$-dimensional hypercube $\Omega = \mathcal{I}_1 \times \cdots \times \mathcal{I}_N$ with $\mathcal{I}_d = [a_d, b_d]$ discretised by uniform grid nodes $(x_{1,i_1}, \ldots, x_{N,i_N})$, where $x_{d,i_d} = a_d + h_di_d$, $i_d = 1, \ldots, n_d$, $n=2^l_d$, $d=1,\ldots,N$.
Let tensor $\bfp$ be its elements 
\begin{align} \label{eq: tensor for multivariate Gaussian}
\bfp_{i_1, \ldots, i_N} = \prod_{d=1}^N \exp \left( - \frac{(x_{d,i_d} - \mu_d)^2} { 2 \sigma_d^2 }\right).
\end{align}
Suppose in addition that $\max_{1 \leq d \leq N} \left( \sqrt{2 \pi} \sigma_d - \int_{a_d}^{b_d} \exp \left( { - \frac{(x_d - \mu_d)^2}{2\sigma_d^2}} \right) \right) \leq \epsilon < 2$.
Then for sufficiently small $\epsilon > 0$, there exists the QTT approximation $\hat{\bfp}$ with ranks bounded as $R_d = 1$ for $d=1,\ldots,N$, and 
\begin{align} \label{eq: QTT rank bounds for multi-dimensional Gaussian}
	R_{d,\ell} \leq C \frac{L_d}{2\sigma_d} \sqrt{\log \left( \frac{1}{\epsilon} \frac{\sigma_d}{1 + L_d/2} \right) },
\end{align}
and the accuracy 
\begin{align} \label{eq: QTT error bounds for multi-dimesnional Gaussian}
	\| \bfp - \hat{\bfp} \| \leq
	\max_d\left( 
	\frac{C}{\sigma_d} \sqrt{\log \left( \frac{1}{\epsilon} \frac{\sigma_d}{1 + L_d/2} \right) } + 1
	\right)N \epsilon,
\end{align}
where $L_d = (b_d - a_d)/2$ and $C$ is a constant does not depend on $L_d$, $\sigma_d$, $\epsilon$, $n_d$, or $N$.
\end{lemma}
\medskip

For fixed $h_d$'s, a more concentrated Gaussian distribution, with a larger ratio of  $L_d$ over $\sigma_d$, would give rise to smaller separation ranks in \eqref{eq: QTT rank bounds for multi-dimensional Gaussian} and smaller approximation error in \eqref{eq: QTT error bounds for multi-dimesnional Gaussian}. 
On the other hand, this could also be achieved by increasing  $h_d$'s while fixing $L_d$'s and $\sigma_d$'s.

\medskip
\begin{remark} \label{remark: QTT representation of multivariate Gaussian}
{\rm
Tolerance $\epsilon$ implicitly impose restrictions on the choice of $L_d$ and $\sigma_d$. By requiring $\exp \left( - \frac{(b_d - \mu_d)^2}{2 \sigma_d^2} \right) = \exp \left( - \frac{(a_d - \mu_d)^2}{2 \sigma_d^2} \right) \leq \epsilon$, we have $L_d \geq 2\sqrt{2} \sigma_d \sqrt{\log \epsilon^{-1}}$, such that $R_{d,\ell} \sim \mathcal{O}(\log(1/\epsilon))$ and $\| \bfp - \hat{\bfp} \| \sim \mathcal{O} (N \epsilon)$. Thus, an estimate for approximation error w.r.t the tensor ranks could be $\| \bfp - \hat{\bfp} \| \sim \mathcal{O} (N \exp(-R))$.
}
\end{remark}

\subsubsection{Non-Gaussian distributions}

In the following, we generalize the QTT approximation of the Gaussian solutions to the cases of more general classes of $N$-dimensional distributions.
Let us consider the class of $\bfp_t$ in  \eqref{eq: TT for vectors} equivalent to certain analytical functions $P_t (\bfx)$ by grid point evaluation.
We assume that $P_t(\bfx)$ allows the efficient approximation in the set of Gaussian distributions on $\Omega$.
Then we prove the following error bound for the QTT approximation.

\medskip
\begin{proposition} \label{theorem: qtt for N-dimensional non-gaussian}
Let $\Omega = \mathcal{I}_1 \times \cdots \times \mathcal{I}_n$ with $\mathcal{I}_d = [a_d, b_d]$, $d=1, \ldots, N$. 
Suppose that for a given continuous function $P: \Omega \rightarrow \mathbb{R}$, and given $\epsilon >0$, there is an approximation by Gaussian sums such that
\begin{align}  \label{eq: assumption for non-gaussian distributions}
	\max_{\bfx\in \Omega} 
	\left| 
	P(\bfx) - \sum_{\ell=1}^Z c_\ell G(\boldsymbol{\mu}^{(\ell)}, \boldsymbol{\sigma}^{(\ell)})
	\right|
	\leq \epsilon,
\end{align}
where $G(\boldsymbol{\mu}^{(\ell)}, \boldsymbol{\sigma}^{(\ell)})$ is the $N$-dimensional Gaussian function defined by $G(\boldsymbol{\mu}^{(\ell)}, \boldsymbol{\sigma}^{(\ell)}) = \prod_{d=1}^N \exp \left( - (x_d - \mu^{(\ell)}_d)^2 / 2 (\sigma^{(\ell)}_d)^2 \right)$. 
In addition, we  assume that $\max_{\substack{1 \leq d \leq N \\ 1 \leq \ell \leq Z}}  \sqrt{2 \pi} \sigma_d^{(\ell)} - \int_{a_d}^{b_d} \exp \left( { - (x_d - \mu^{(\ell)}_d)^2 / 2 \left( \sigma^{(\ell)}_d \right)^2}  \right) \leq \epsilon < 2$.
Then, consider an $N$-dimensional tensor $\bfp$ defined by its entries
$\bfp_{i_1, \ldots, i_N} = P(x_{1,i_1}, \ldots, x_{N,i_N})$, 
for $i_d = 1, \ldots, n_d$, $d= 1, \ldots, N$,
where $x_{d,i_d} = a_d + h_di_d$, $h_d = (b_d - a_d) / n_d$.
It allows an QTT tensor $\hat{\bfp}$ in the form of \eqref{eq: TT for vectors}, with ranks bounded by $R_d = Z$  and 
\begin{align}  \label{eq: QTT rank bounds for multi-dimensional non-Gaussian}
	R_{d,\ell} \leq C_1 \frac{Z L_d}{ \min_\ell\sigma_d^{(\ell)}} \sqrt{\log \left( \frac{1}{\epsilon} \frac{\max_\ell \sigma_d^{(\ell)}}{1+L_d/2} \right)},
\end{align}
for $\ell = 1, \ldots, n_d$, $d=1, \ldots, N$,
and the accuracy
\begin{align*}
	\| \bfp - \hat{\bfp} \| \leq
	C_2ZN \epsilon,
\end{align*}
where $C_1$ is a constant does not depend on $L_d$, $\sigma_d^{(\ell)}$, $\epsilon$,  $n_d$, $N$, or $Z$,
and $C_2$ is a constant does not depend on $N$, $Z$, or $\epsilon$.
\end{proposition}
\medskip

\begin{proof}
	We define tensors $\mathbf{g}_\ell$ by grid evaluation of the $N$-dimensional Gaussian $G(\boldsymbol{\mu}^{(\ell)}, \boldsymbol{\sigma}^{(\ell)})$, for $\ell = 1, \ldots, Z$, as in \eqref{eq: tensor for multivariate Gaussian}. From Lemma \ref{lemma: qtt for N-dimensional gaussian}, there exists an QTT approximation $\hat{\mathbf{g}}_\ell$, with ranks $R^{(\ell)}$ bounded by \eqref{eq: QTT rank bounds for multi-dimensional Gaussian} and accuracy given by \eqref{eq: QTT error bounds for multi-dimesnional Gaussian}.
Hence, we define $\hat{\bfp}$, as a QTT approximation of $\bfp$, by $\hat{\bfp} = \sum_{\ell=1}^Z \hat{\mathbf{g}}$.
Using the addition rules of QTT ranks~\cite{kazeev2012low}, the bounds in \eqref{eq: QTT rank bounds for multi-dimensional non-Gaussian} can be justified.

To prove the accuracy, we use the triangular inequality
\begin{align*}
	\| \bfp - \hat{\bfp} \| 
	\leq 
	\| \bfp - \sum_{\ell=1}^Z \mathbf{g}_\ell \| + \sum_{\ell=1}^Z \| \mathbf{g}_\ell - \hat{\mathbf{g}}_\ell \|,
\end{align*}
where the first term is bounded by the assumption \eqref{eq: assumption for non-gaussian distributions}, and a bound from the second term can be inferred from Remark \ref{remark: QTT representation of multivariate Gaussian}.
\end{proof}

\medskip
\begin{remark} \label{remark: QTT representation of multivariate non-Gaussian}
{\rm 
Similarly as in Remark \ref{remark: QTT representation of multivariate Gaussian}, we can derive, from Proposition \ref{theorem: qtt for N-dimensional non-gaussian}, that $R_{d,\ell} \sim \mathcal{O}(Z \log(1/\epsilon))$, and $\| \bfp - \hat{\bfp} \| \sim \mathcal{O}(ZN \epsilon)$. 
Now, if we define tolerance $\varepsilon$ by requiring $\| \bfp_t - \hat{\bfp}_t \| \sim \mathcal{O} (\varepsilon)$, where $\bfp$ and $\hat{\bfp}_t$ are defined in \eqref{eq: CFPE eigenvalue problem after tensor truncation} and \eqref{eq: TT for vectors}, respectively,
then there exists a QTT representation $\hat{\bfp}_t$ whose separation ranks scale as 
\begin{align}
R \sim \mathcal{O}(Z \log(ZN/ \varepsilon)),
\end{align}
where $R \geq R_d, R_{d,\ell}$, $\ell = 1, \ldots, l_d$, $d = 1, \ldots, N$.
}
\end{remark}
\medskip

Back to our question at the beginning of this section about the existence of a low-rank $\epsilon$-approximation to the solution of \eqref{eq: CFPE eigenvalue problem after tensor truncation}. 
Remark \ref{remark: QTT representation of multivariate non-Gaussian} imposes a key condition that the eigenfunctions need to be well approximated by the sum of a minimum number of Gaussian functions. 
And the peaks of these Gaussian functions have to be significantly away from the boundary $\partial \Omega$. Unfortunately, conditions on the operator $\mathbf{A}_t$ are still unclear.

\subsection{Higher order inverse iteration}

Remark \ref{remark: QTT representation of multivariate non-Gaussian} ensures that one class of the eigenvector $\bfp_t$ allows a low rank QTT $\varepsilon$-approximation  as in Proposition \ref{theorem: qtt for N-dimensional non-gaussian}.
To approximate the eigenpair $(\lambda_t, \bfp_t)$, we use a higher order analogue of the inverse iteration, combined with tensor truncations. 
The main building block is as follows.

\begin{algorithm}[Inverse power method in tensor format] \label{algorithm: inverse power method}
\\ \indent \indent For $k=1, 2, \ldots$ till convergence do
\\ \indent \indent \indent $(\mathbf{A}_t - \shift \mathbf{I} ) \bfp_{k} = \bfp_{k-1} + \mathbf{r}_k$, with $\| \mathbf{r}_k \| \leq \varepsilon \| \bfp_k \|$;
\\ \indent \indent \indent $\bfp_k = \bfp_k / \| \bfp_k \|$;
\\ \indent \indent end
\end{algorithm}

When the linear systems are solved precisely, i.e., $\| \mathbf{r}_k \| \equiv 0$, Algorithm \ref{algorithm: inverse power method}, beginning with an initial tensor $\bfp_0$, the series $\{ \bfp_{k} \} _{k = 1, 2, \ldots }$
would converge to the eigenvector corresponding to the eigenvalue closest to the chosen shift $\shift$. 
If we assume the perturbation $\delta \mathbf{A}_h$ in \eqref{eq: CFPE eigenvalue problem after tensor truncation} is sufficiently small that the matrix corresponding to $\mathbf{A}_t$ remains $M$-matrix, then from Lemma \ref{lemma: A_h => M-matrix}, any non-negative $\shift$ would lead to the correct convergence direction towards $(\lambda_t, \bfp_t)$.

However, there are two main reasons that the residue tensor $\mathbf{r}_k$ needs to be considered. First, we use the Alternating minimal energy method (AMEN)~\cite{dolgov2013alternating1,dolgov2013alternating2} to conduct inner iterations to solve the linear system in QTT format in Algorithm \ref{algorithm: inverse power method}. A highly accurate solution $\bfp_k$ in each inverse iteration requests more computational time, and it is usually not necessary (as we will prove later).
Second, tensor rounding procedure needs to be performed after each inverse iteration to avoid uncontrollable growth of the tensor separation rank~\cite{oseledets2011tensor,holtz2012the}, which also adds to the residual.
Therefore, it is of interest to analyse the effect of the residues $\mathbf{r}_k$ on the final convergence.

\subsection{Algebraic error}

The error analysis of Algorithm \ref{algorithm: inverse power method} is analogous to the analysis for the inexact inverse power method \cite{golub2000inexact}.
Let $(\lambda_t^i,\mathbf{u}_i,\mathbf{v}_i)$ for $i = 1,2,\ldots,n$ be a complete set of eigentriples of $\mathbf{A}_t$ satisfying $0 >\lambda_t^1 > \lambda_t^2 \geq \cdots $. It follows that
\begin{align*}
	\mathbf{u}_i^{T} \mathbf{A}_t = \lambda_t^i \mathbf{u}_i^{T} , \ \ \mathbf{A}_t \mathbf{v}_i = \lambda_t^i  \mathbf{v}_i \ \ \mathrm{and} \ \ \mathbf{u}_i^T \mathbf{A}_t \mathbf{v}_j = \delta_{i,j}, \ \ \mathrm{for} \ \ i,j = 1,2,\ldots,n,
\end{align*}
where $\delta_{i,j}$ is the Kronecker symbol. 
We assume that the intermediate solution after $k$ inverse iterations can be expanded as a linear combination of right eigenvectors:
\begin{align} \label{eq: expansion of p_k}
	\bfp_k = \sum_{i=1}^n c^{(k)}_i \mathbf{v}_i, \  \ \mathrm{where} \  \ c_i^{(k)} = \mathbf{u}_i^T \mathbb{A}_t \bfp_k.
\end{align}
Now we define a measure of the approximation of $\bfp_k$ to $\mathbf{v}_1$ for $c_1^{(k)} \neq 0$ as
\begin{align*}
t_k = \| ( c_2^{(k)},c_3^{(k)},\ldots,c_n^{(k)} ) \| / | c_1^{(k)} |.
\end{align*}
The convergence of $t_k$ is concluded by the corresponding matrix analysis, see Lemma 2 and 3 in \cite{golub2000inexact}, and for Algorithm \ref{algorithm: inverse power method}, we state the tensor version as below.

\medskip
\begin{lemma} \label{lemma: converge of inverse iteration}
	Let $\rho = \left| (\lambda_1  - \shift) / (\lambda_2 - \shift) \right| <1$, and $c^{(k)}_1 \neq 0$. Let columns of tensor matrices $\mathbf{U}$ and $\mathbf{V}$ contain all left and right eigenvectors of $\mathbf{A}_t$, respectively. Then,
	\begin{align} \label{eq: converge of inverse iteration in t}
		t_k \leq \rho t_{k-1} + \varepsilon C,
	\end{align}
where $C \leq \| \mathbf{U}^T \| \| \mathbf{V}^T \| (1 + t_0)^2$.
\end{lemma}
\medskip

Then we have the convergence of the algebraic error in the following theorem.

\medskip
\begin{theorem} \label{theorem: algebiraic error convergence}
	Consider $\sigma > 0$ in Algorithm \ref{algorithm: inverse power method}, and the spectrum of $\mathbf{A}_t$ is distributed in the negative half plane. Let $\langle \bfp_t, \bfp_k \rangle \neq 0$ and let $\bfp_k$ be normalised such that $\langle \bfp_t, \bfp_k \rangle = 1$. Then, the convergence of $\bfp_k$ towards the principal eigenvector $\bfp_t$ of $\mathbf{A}_t$ is given by 
\begin{align} \label{eq: algebraic error convergence}
	\| \bfp_t - \bfp_k \| \leq C_1 \rho^k + \varepsilon C_2 \frac{1 - \rho^k}{1 - \rho},
\end{align}
where $\rho$ is defined in Lemma \ref{lemma: converge of inverse iteration}, and $C_1$ and $C_2$ are constants that do not depend on $\rho$ and $k$.
\end{theorem}
\medskip

\begin{proof}
It follows directly from \eqref{eq: converge of inverse iteration in t} that, for $\forall k =1,2,\dots$, we have
\begin{align*}
	t_k 
	& \leq  \rho t_{k-1} + \varepsilon C 
	\leq  
	\cdots 
	\leq \rho^k t_{0} + \rho^{k-1} \varepsilon C + \cdots +  \varepsilon C
	= \rho^k t_{0} + \varepsilon C \frac{1 - \rho^k}{1 - \rho}.
\end{align*} 
By requesting $\langle \bfp_t, \bfp_k \rangle = 1$, we have $c_1^{(k)} = 1$. Then,
\begin{align*}
	\| \bfp_t - \bfp_k \| = \| \sum_{i=2}^n c_i^{(k)} \mathbf{v}_i \| \leq  \sum_{i=2}^n c_i^{(k)} \| \mathbf{v}_i \| \leq t_k,
\end{align*}
where $c_i^{(k)}$ and $\mathbf{v}_i$ are defined in \eqref{eq: expansion of p_k}.
\end{proof}

\medskip
\begin{remark} \label{remark: inexactness of the inverse power method}
{\rm 
The convergence rate in \eqref{eq: algebraic error convergence} suggests that, for $\rho$ small, Algorithm \ref{algorithm: inverse power method} would still give convergence with algebraic error of $\mathcal{O}(\varepsilon)$. The inexactness of the tensor linear solvers and the tensor rounding procedure will dominate the algebraic error in the final stage of computation.
}
\end{remark}

\section{Numerical illustrations}

We run all our numerical experiments in Matlab solely on a MacBook Pro laptop (OS X 10.9.5) with a 2 GHz Intel Core i7 processor and 8 GB of physical memory.
Our source codes made extensive use of the Tensor Train toolbox~\cite{oseledets2011tensor}, and is part of the Stochastic Bifurcation Analyzer toolbox freely available at \texttt{http://people.maths.ox.ac.uk/liao/stobifan/index.html} \cite{liao2014parameter}.

\subsection{1-D birth-death process}
As a first application of our theory, we will estimate all sources of errors discussed in sections 2-6 for the birth-death process.
This is the simplest case of a molecular reaction mechanism.
The main purpose of considering such a reaction is that both its stationary CME \eqref{eq: CME} and CFPE \eqref{eq: CFPE} are exactly solvable and hence it provides us with a direct test of our expressions in the error estimates.
The set of reactions under study are
\begin{align} \label{eq: birth-death reactions}
	\mbox{ \raise 0.851 mm \hbox{$\emptyset$}}
\;
\mathop{\stackrel{\displaystyle\longrightarrow}\longleftarrow}^{k_1}_{k_{2}}
\;
\mbox{\raise 0.851 mm\hbox{$X$}}.
\end{align}
A single chemical species, denoted as $X$, is produced by some substrates within certain container of volume $\Vol$ at a constant rate $k_1$, and de-gradates with rate constant $k_2$.

The CME \eqref{eq: CME} for the birth-death reactions \eqref{eq: birth-death reactions} reads
\begin{align} \label{eq: birth-death CME}
	\alp_1 (n+1) \Pms (n+1) + \alp_2 (n-1) \Pms (n-1) - \alp_1 (n) \Pms (n) - \alp_2 (n) \Pms (n) = 0,
\end{align}
for $n \in \mathbb{N}$, where the propensities functions are
\begin{align*}
	\alp_1 (n) = k_1 \Vol \ \ \mathrm{and} \ \ \alp_2 (n) = k_2 n.
\end{align*}
The stationary solution $\Pms (n) $ of \eqref{eq: birth-death CME} is the Poisson distribution 
\begin{align} \label{eq: birth-death CME solution}
	\Pms (n) = \frac{1}{n!} \left( \frac{k_1 \Vol}{k_2} \right)^n \exp \left[ - \frac{k_1 \Vol}{k_2} \right].
\end{align}
The corresponding Fokker-Planck approximation \eqref{eq: CFPE} of the CME \eqref{eq: birth-death CME} can be written as
\begin{align} \label{eq: birth-death CFPE}
	\frac{1}{2} \frac{\pd ^2}{\pd x^2} \left[ (k_1 \Vol  + k_2 x) \Pfs (x) \right]
	- \frac{\pd }{\pd x} \left[ (k_1 \Vol - k_2 x) \Pfs (x) \right] = 0,
\end{align}
for $x \in \D_\infty = (-k_1 \Vol / k_2, \infty)$.
Integrating over $x$ and using the boundary conditions $\Pfs (x)  \rightarrow 0$ as $x  \rightarrow + \infty$, we obtain 
\begin{align} \label{eq: birth-death CFPE solution}
	\Pfs (x) = 2C \exp \left[ -2 x + \left( \frac{4k_1 \Vol}{k_2} - 1 \right)  \log (k_1\Vol + k_2 x) \right],
\end{align}
where the normalisation constant $C$ is chosen such that $\int_{\D_\infty} \Pfs (x) \pd x = 1$, i.e. 
\begin{align*}
	C = \left( \int_{\D_\infty} \exp \left[ -2x + \left( \frac{4k_1 \Vol}{k_2} - 1 \right) \log (k_1 \Vol + k_2 x ) \right]  \right)^{-1}.
\end{align*}

\begin{figure}[ht] 
\noindent
\rule{0pt}{0pt} 
\raise 5.2cm \hbox{(a)}
\hskip -7mm
\includegraphics[scale=0.34]{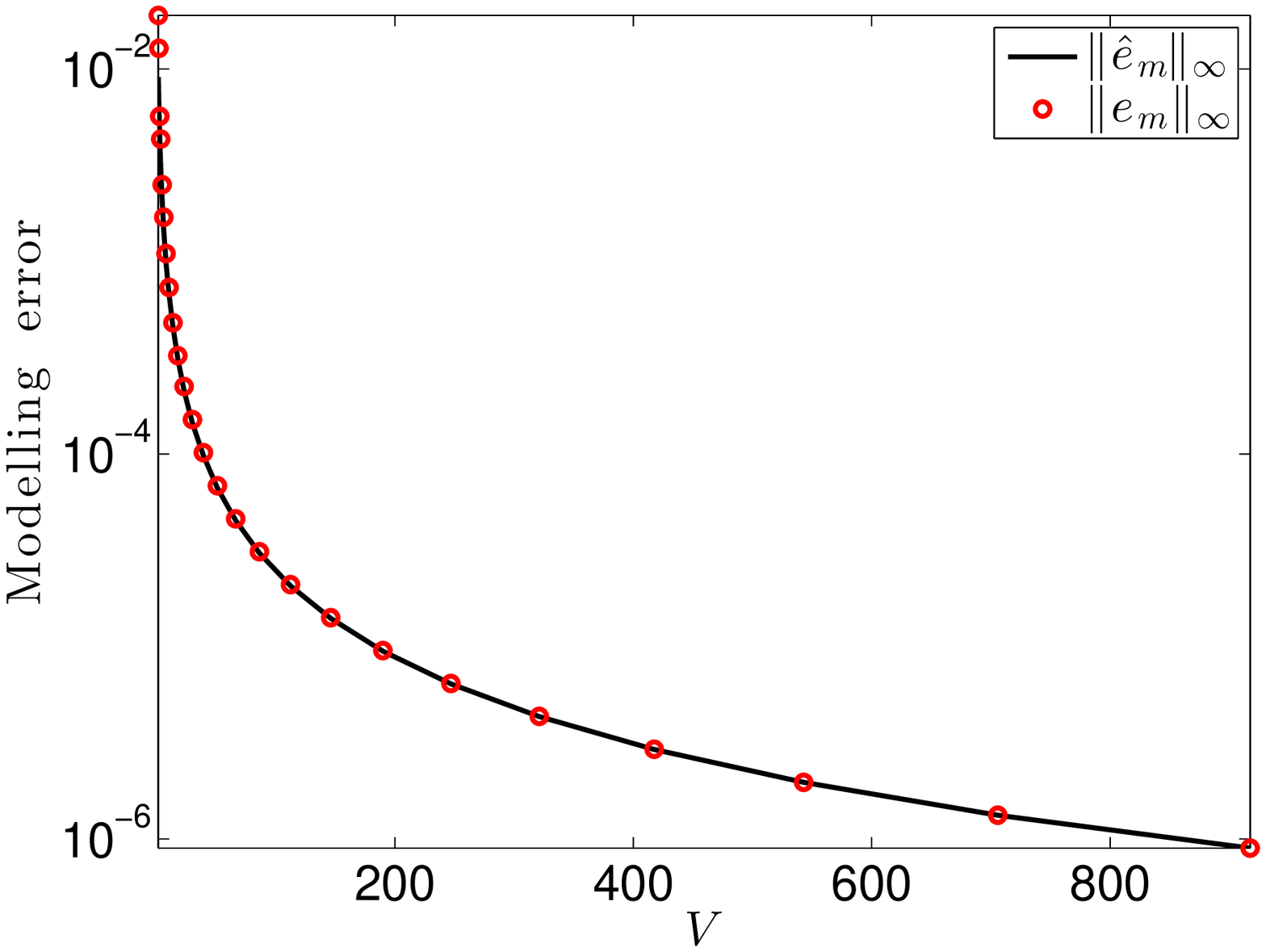}
\hskip 0mm
\raise 5.2cm \hbox{(b)}
\hskip -6mm
\includegraphics[scale=0.34]{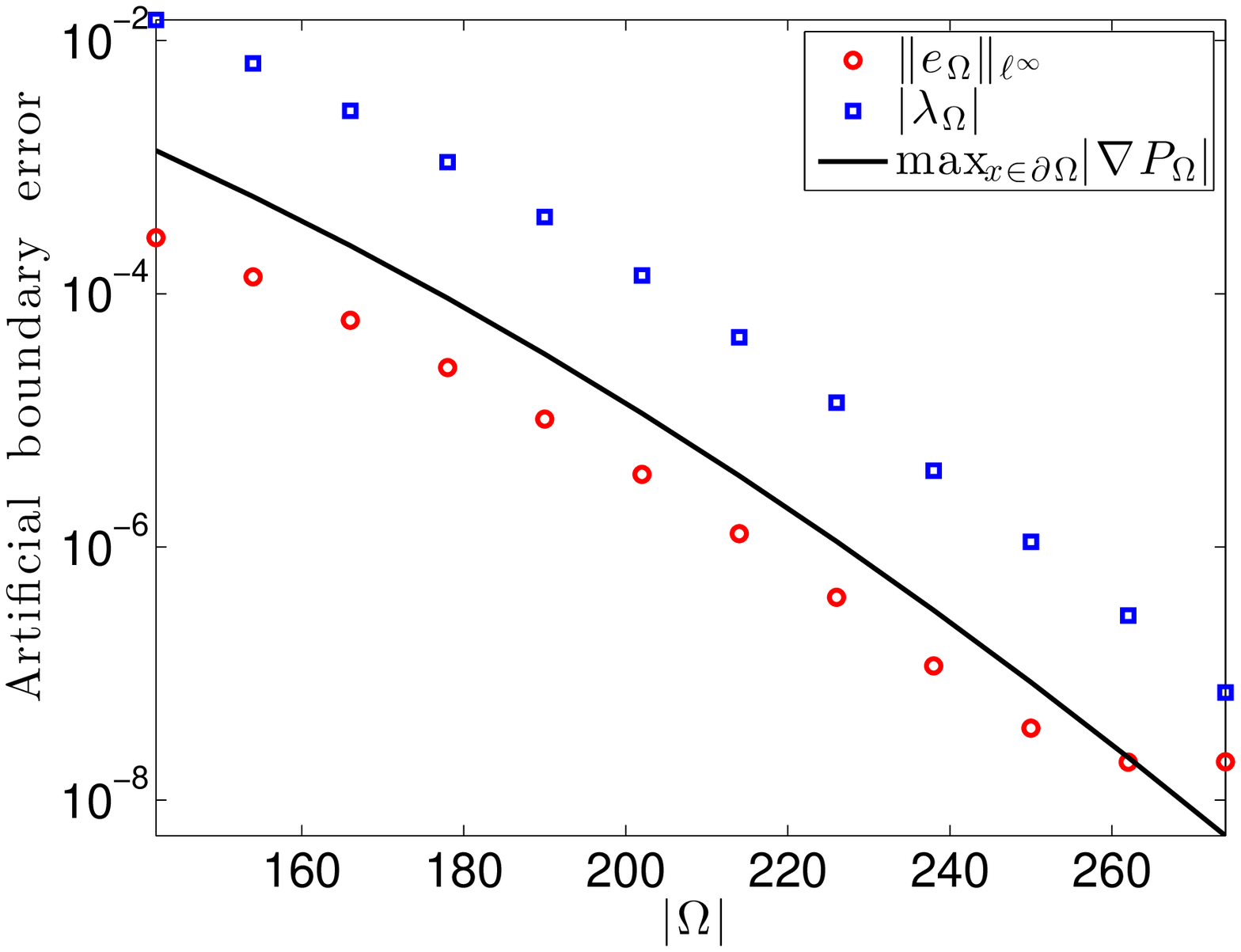}

\vskip 0.5mm

\hskip -0mm
\raise 5.2cm \hbox{(c)}
\hskip -7mm
\includegraphics[scale=0.34]{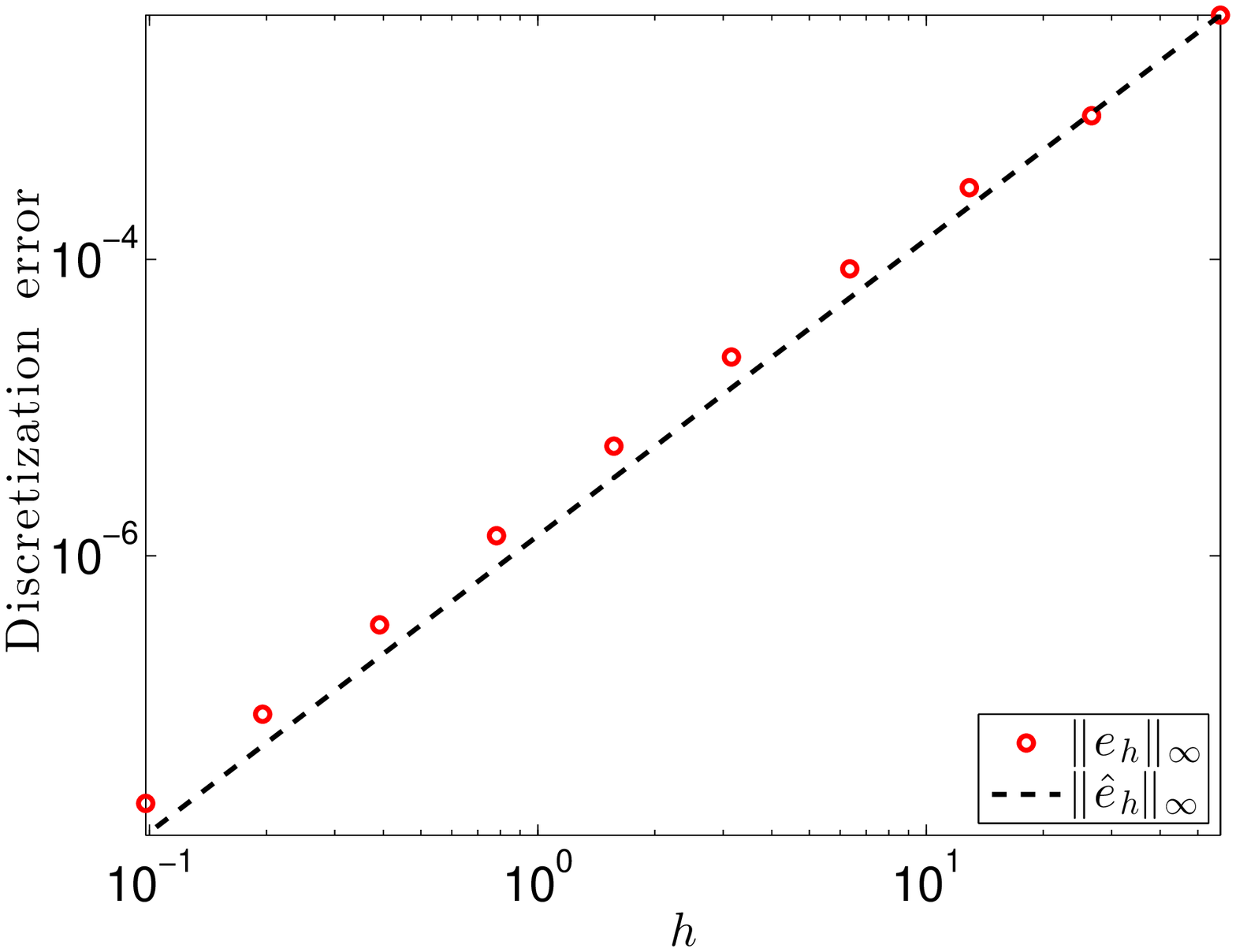}
\hskip 0mm
\raise 5.2cm \hbox{(d)}
\hskip -8mm
\includegraphics[scale=0.34]{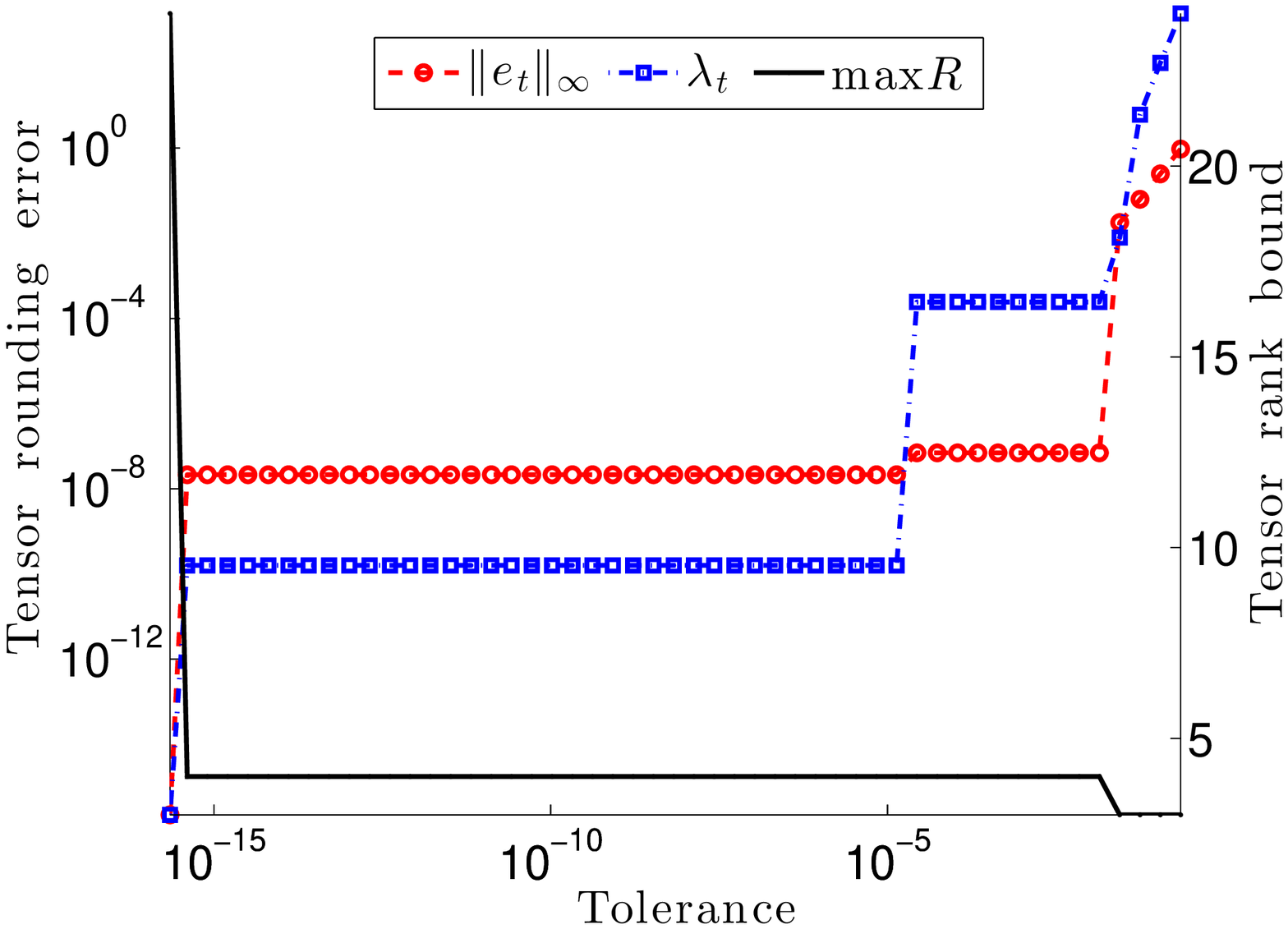}
\caption{ \label{fig: birth-death}
Application of the error analysis to the birth-death process \eqref{eq: birth-death reactions}. If not specified, $k_1 = k_2 = 1$, $\Vol = 500$ and $\D = [300, 700]$.
{\rm (a)} 
Modelling error of the CFPE approximation \eqref{eq: birth-death CFPE} of the CME \eqref{eq: birth-death CME}.
The red circles represents the exact difference between the analytic solutions given in \eqref{eq: birth-death CME solution} and \eqref{eq: birth-death CFPE solution}.
The black solid curve refers to the right hand side of \eqref{eq: modellingerrdetailbalance} with fitted parameter values: $C = 1.43 \times 10^{-5}$.
{\rm (b)} 
The domain size $|\D|$ dependence of the artificial boundary error in \eqref{eq: birth-death CFPE Omega} with $\Vol = 500$.
Error $e_\D$ is defined as $\Pfs |_{x \in \D} - \Pfb$, where $\Pf$ is given in \eqref{eq: birth-death CFPE solution} and $\Pfb$ is approximated here by a finite difference solution with grid size $h=0.1$.
{\rm (c)} 
Discretization error of the monotone difference scheme \eqref{eq: finite difference scheme} in solving \eqref{eq: birth-death CFPE Omega}.
The exact error is plotted with red circles, and the dashed curve refers to the reference line $C \cdot h^2$ with $C=1.37 \times 10^{-6}$.
{\rm (d)}
Left y-axis: The principal eigenvalue $\lambda_t$ and the tensor rounding error, $e_t \equiv \bfp_h - \bfp_t$, where $\bfp_h$ is defined in \eqref{eq: canonical tensor for A_h},  and $(\lambda_t, \bfp_t)$ are defined  in \eqref{eq: CFPE eigenvalue problem after tensor truncation}. Right y-axis: Maximum QTT ranks of the assembled tensor matrix $\mathbf{A}_t$. The lower bound for x-axis is the machine epsilon $2.2204 \times 10^{-16}$, meaning that no tensor rounding was performed at that point. The computational domain $\D$ is discretised by $2^{10}$ equidistant nodes.
}
\end{figure}

With explicit formulas \eqref{eq: birth-death CME solution} and \eqref{eq: birth-death CFPE solution}, we evaluate the exact modelling error $e_m \equiv \Pms - \Pfs$, and plot the error measured in $\ell^\infty$-norm in Fig. \ref{fig: birth-death}(a) against increasing values of system volume $\Vol$. 
As comparison, we use the black curve to refer to our estimate $\| \hat{e}_m \|_\infty$ in \eqref{eq: modellingerrdetailbalance} of Theorem \ref{theorem: modellingerrdetailbalance}.
By fitting the constant coefficients in \eqref{eq: modellingerrdetailbalance} to the exact errors $\| e_m \|_\infty$, we obtain a good agreement between the exact errors and the estimated ones.

Next, we consider approximating the CFPE \eqref{eq: birth-death CFPE} within a bounded domain $\D \subset \D_\infty$. 
As discussed in section 3, the stationary solution $\Pfs$ in \eqref{eq: birth-death CFPE solution} is approximated by the positive principal eigenfunction $\Pfb$, satisfying
\begin{align} \label{eq: birth-death CFPE Omega}
\begin{cases}
	\frac{1}{2} \frac{\pd ^2}{\pd x^2} \left[ (k_1 \Vol  + k_2 x) \Pfb (x) \right]
	- \frac{\pd }{\pd x} \left[ (k_1 \Vol - k_2 x) \Pfb (x) \right] = \lambda_\D \Pfb (x),  & x \in \D,
	\\
	\Pfb (x) = 0,  &x \in \partial \D, 
\end{cases}	
\end{align}
where $\lambda_\D$ is the principal eigenvalue.
In Fig. \ref{fig: birth-death}(b), we show the domain $\D$ dependence of $\lambda_\D$ and $e_\D \equiv \Pfs |_{x \in \D } - \Pfb$. 
Here, we fix the center of $\D$ at the mean value $x=500$, and vary the domain size $|\D|$.
In accordance with our convergence statement in Theorem \ref{theorem: artificial boundary condition}, the numerical experiment also shows that $\lambda_\D \rightarrow 0$ and $\| e_\D \| _\infty \rightarrow 0$ as $| \D |$ increases.
We also plot the changes of  $\max_{x \in \partial \D} |\nabla \Pfb|$ with respect to $| \D |$ as the solid cure in Fig. \ref{fig: birth-death}(b).
Such quantity represents the maximum gradient of the solution at the boundaries. We can observe that the decay rate of $\lambda_\D$ is similar to $\| e_\D \| _\infty$, indicating that both these quantities can be potentially used as error indicators for the artificial boundary error.
Unfortunately, we are not able to provide theoretical proof for such argument at the moment.

Fig. \ref{fig: birth-death}(c) illustrates the grid-size $h$ dependence of the discretisation error in numerical solution of \eqref{eq: birth-death CFPE Omega} using the difference scheme \eqref{eq: finite difference scheme} in section 4.
The discrete principal eigenfunction $p_h$ is computed using Matlab \texttt{eigs} function.
The exact discretization error, $e_h = \Pfb - p_h$, is then approximately computed by $e_h \approx e_h + e_\D$ by choosing $| \Omega | = 400$ such that $e_\D = \Pfs | _{x \in \D} - \Pfb$ is sufficiently small.
We could observe that the convergence of the discretization error $\| e_h \| _\infty$ (red circles) agrees well with our error estimate $\| \hat{e}_h \|_\infty \equiv C h^2 $ (dashed curve) derived in \eqref{theorem fdm convergence} of Theorem \ref{eq: theorem fdm eigenvalue convergence}.

Following sections 5.1, 5.2 and 5.3, we now establish the QTT representation \eqref{eq: qtt} of the difference scheme \eqref{eq: finite difference scheme} for the eigenvalue problem \eqref{eq: birth-death CFPE Omega}, and test how the tensor rounding procedure of the assembled tensor matrix would cause deflection in the principal eigenpair $(\lambda_h, p_h)$ towards $(\lambda_t, \bfp_t)$.
Since the error in the rounding of the tensor matrix is of the main concern, for different tolerances in tensor rounding, we always first assemble the operator $\mathbf{A}_h$ in QTT format, apply TT-rounding algorithm~\cite{oseledets2011tensor} with the prescribed tolerance to truncate $\mathbf{A}_h$ to form $\mathbf{A}_t$, unfold tensor matrix $\mathbf{A}_t$ into the standard matrix $A_t$, and then use the Matlab \texttt{eigs} function to compute the principal eigenpair $(\lambda_t, p_t)$.
Here, $p_t$ is the equivalent vector form of the tensor $\bfp_t$.
In this way, we could exactly measure the error $e_t = p_h - p_t$ caused by tensor rounding of the QTT matrix $\mathbf{A}_h$. 
In Fig. \ref{fig: birth-death}(d), we could see the trend that the error $e_t$ in $\ell^\infty$-norm increases for larger tolerance values.
But we could also observe that, for ranges of tolerance values, the tensor rounding error $\| e_t \|_\infty$, together with $\lambda_t$, remains at a constant level. 
This is not predicted by our Theorem \ref{theorem: tensor truncation error}.
The reason is that the tensor rounding algorithm is SVD-based~\cite{oseledets2011tensor}, meaning that the rank truncation is based on the magnitude of the singular values rather than certain matrix norm.
Therefore, it is possible that different tolerance values in tensor rounding procedure would generate the same result, especially when there exist large spectrum gaps.
Also, as predicted by Theorem \ref{theorem: CFPE ranks}, the maximum QTT rank \eqref{eq: qtt} of $\mathbf{A}_h$ is 24 (see Fig. \ref{fig: birth-death}(d)).
But even for very small tolerance values, the maximum rank of $\mathbf{A}_t$ is reduced to 4, by only causing an error of order $10^{-8}$ in the approximated principal eigenvector.

\begin{figure}[t] 
\noindent
\rule{0pt}{0pt} 
\raise 5.2cm \hbox{(a)}
\hskip -7mm
\includegraphics[scale=0.34]{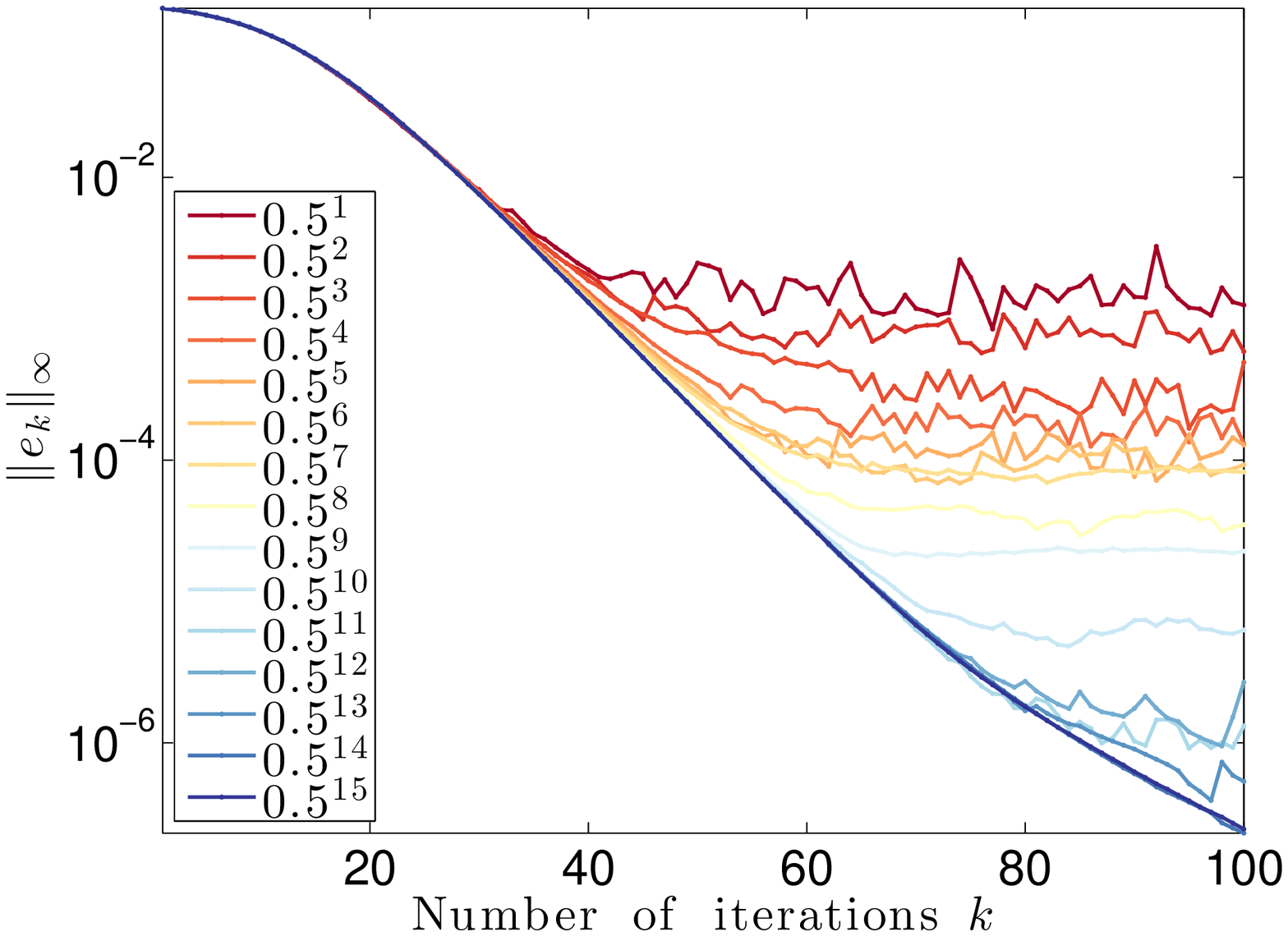}
\hskip 0mm
\raise 5.2cm \hbox{(b)}
\hskip -8mm
\includegraphics[scale=0.34]{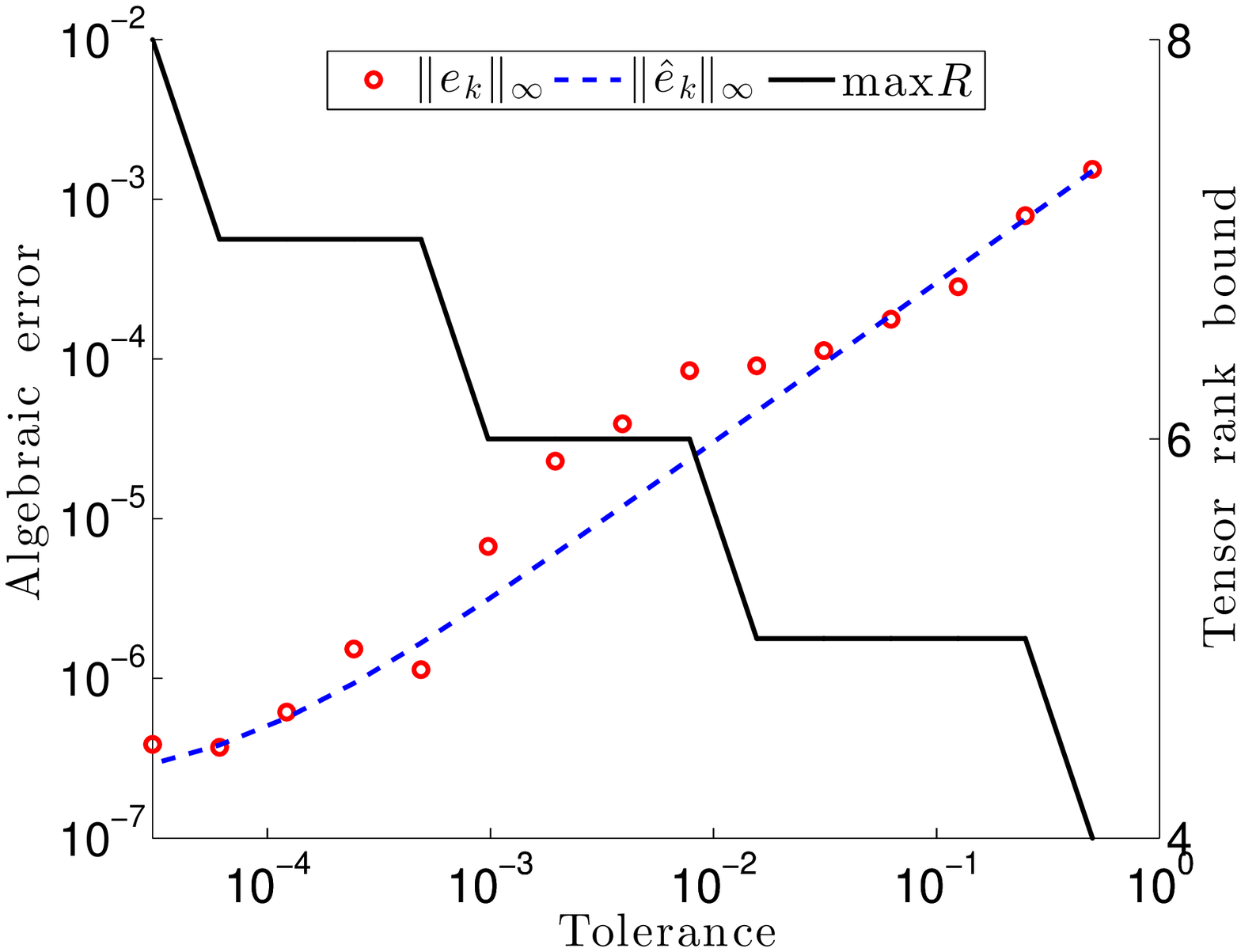}
\caption{ \label{fig: birth-death algebraic}
Algebraic error analysis of the tensor method applied to the birth-death process \eqref{eq: birth-death reactions}. Parameters: $k_1 = k_2 = 1$, $\Vol = 500$, $\D = [300, 700]$, $n=2^{10}$, $\varepsilon_t=10^{-10}$, $\sigma = 10$.
{\rm (a)} 
Algebraic error in $\ell^\infty$-norm for the first 100 inverse iterations in Algorithm \ref{algorithm: inverse power method}. The tensor linear systems are solved using the AMEN under the tolerance values given in the legend.
{\rm (b)}
Left y-axis: Algebraic error after $100$ inverse iterations. 
The red circles mark the exact error, and the dashed curve refer to $y=C_1 x + C_2$ with $C_1 = 0.003$ and $C_2 = 2 \times 10^{-7}$.
Right y-axis: The maximum QTT ranks in the final solution.
}
\end{figure}

Next, instead of using the Matlab \texttt{eigs} function, we keep all objects in QTT format and solve for $\bfp_t$ using the higher order inverse iteration presented in Algorithm~\ref{algorithm: inverse power method}.
Setting the shift value $\sigma = 0.1$, we computed the algebraic error $e_k \equiv \bfp_t - \bfp_k$ for each of the $100$ inverse iterations. The convergences of the algebraic error $e_k$ in $\ell^\infty$-norm under difference stopping tolerances for the tensor linear solver -- AMEN~\cite{dolgov2013alternating1, dolgov2013alternating2} are shown in Fig. \ref{fig: birth-death algebraic}(a).
We see that all simulations initially converge with similar rates, and the convergence comes to a halt after certain number of iterations.
This matches our prediction in Theorem \ref{theorem: algebiraic error convergence}, that for small number of iterations $k$, the term $C_1 \rho^k$ in \eqref{eq: algebraic error convergence} dominates the algebraic error $\| e_k \|$, while for large $k$, the dominance is taken over by the term $\varepsilon C_2 \frac{1-\rho^k}{1-\rho}$ in \eqref{eq: algebraic error convergence} caused by the inexactness of the tensor linear solver (Remark \ref{remark: inexactness of the inverse power method}).
In Fig.  \ref{fig: birth-death algebraic}(b), we only consider the algebraic error $\| e_k \|_\infty$ after $k=100$ inverse iterations, and plot it against different tolerance in solving the tensor linear systems. 
The algebraic error agrees well with the reference line $y=C_1x+C_2$ in the blue dashed curve, which is the direct result from \eqref{eq: algebraic error convergence} in Theorem \ref{theorem: algebiraic error convergence} for fixed $\rho$ and $k$.
We could also observe the increase in the tensor rank as the tolerance value decreases (solid curve in Fig. \ref{fig: birth-death algebraic}(b)), and the exponential scaling matches our prediction in Remark \ref{remark: QTT representation of multivariate non-Gaussian}.

\subsection{A 50-dimensional reversible isomerization reaction chain} 

In this section, the simple birth-death process \eqref{eq: birth-death reactions} is extended to a 50-dimensional reaction chain, where molecules are allowed to transform themselves into different isometric forms, i.e.,
\begin{align} \label{eq: isometrization reactions}
	\mbox{ \raise 0.851 mm \hbox{$\emptyset$}}
\;
\mathop{\stackrel{\displaystyle\longrightarrow}\longleftarrow}^{k_1}_{k_{2}}
\;
\mbox{\raise 0.851 mm\hbox{$X_1$}}
\;
\mathop{\stackrel{\displaystyle\longrightarrow}\longleftarrow}^{k_3}_{k_{4}}
\;
\mbox{\raise 0.851 mm\hbox{$X_2$}}
\;
\mathop{\stackrel{\displaystyle\longrightarrow}\longleftarrow}^{k_5}_{k_{6}}
\;
\mbox{\raise 0.851 mm\hbox{$\cdots$}}
\;
\mathop{\stackrel{\displaystyle\longrightarrow}\longleftarrow}^{k_{97}}_{k_{98}}
\;
\mbox{\raise 0.851 mm\hbox{$X_{49}$}}
\;
\mathop{\stackrel{\displaystyle\longrightarrow}\longleftarrow}^{k_{99}}_{k_{100}}
\;
\mbox{\raise 0.851 mm\hbox{$X_{50}$}}
\;
\mathop{\stackrel{\displaystyle\longrightarrow}\longleftarrow}^{k_{101}}_{k_{102}}
\;
\mbox{\raise 0.851 mm\hbox{$\emptyset,$}}
\end{align}
where $X_i$, $i=1, 2, \ldots, 50$, can be interpreted as the $i$-th isometric form of species $X$, and $k_{ j}$, $j=1, 2, \ldots, 102$, are reaction constants.
The corresponding stoichiometric matrix $\boldsymbol{\Vs} \in \mathbb{Z}^{102 \times 50}$ is given by 
\begin{align*}
	\Vs_{j,i} = 
	\begin{cases}
	1 & j= 2i-1 \ \  \mathrm{or} \ \  j=2i+2, \\
	-1 & j= 2i \ \ \mathrm{or} \ \ j=2i+1, \\
	0 & \mathrm{otherwise},
	\end{cases}	
	\quad \mathrm{for} \quad i= 1,\ldots, 50 \; \mathrm{and} \; j=1, \ldots, 102.
\end{align*}
Let $\bfx = [x_1, x_2, \ldots, x_{50}]$ be the state vector, then the propensity functions are of the form 
\begin{align*}
	\alp_j (\bfx) = 
	\begin{cases}
	k_j \Vol & j=1 \; \mathrm{or} \; 202, \\
	k_j x_{j/2} & j \ \mathrm{is} \  \mathrm{even} \\
	k_j x_{(j-1)/2} & \mathrm{otherwise},
	\end{cases}
	\quad \mathrm{for} \quad j= 1, \ldots, 102.
\end{align*}
Consider $\Pms (\bfn)$, $\bfn \in \mathbb{N}^{50}$, be the solution of the stationary CME \eqref{eq: CME} for the isometrization reaction \eqref{eq: isometrization reactions}, it is a product Poisson distribution~\cite{jahnke2007solving},
\begin{align} \label{eq: isomerization Poisson}
	\Pms (\bfn) = \prod_{i=1}^{50} \Pms^i (n_i) 
	= \prod_{i=1}^{50} \frac{\phi_i^{n_i} }{n_i !} e^{- \phi_i},
\end{align}
where $\Pms^i$, $i=1,\ldots, 50$, stand for univariate Poisson functions whose mean $\phi_i$ equal to the stead state solution of a system of reaction rate equations.

In comparison, the analytical formula for the solution of the stationary CFPE \eqref{eq: CFPE} for the isomerization reactions \eqref{eq: isometrization reactions} is not obvious.
Furthermore, because of the very high dimensionality, we would not be able to identify individual sources of errors separately as we did in section 7.1.
This is to say that we would only be able to compute the final tensor approximation $\bfp_k$, whose accuracy depends on the accuracies of all intermediate steps {\bf (a1)--(a1)} in the TPA (Table \ref{tab: TPA}). 
Whereas there will be no intermediate solutions available to guide us to choose appropriate parameters in each step.
To address this issue, we notice that the isomerization reaction chain \eqref{eq: isometrization reactions} could be viewed as an extension of the birth death process \eqref{eq: birth-death reactions}, and thus our error analysis in section 7.1 could potentially hint the choice of parameters for simulating the extended reaction chain \eqref{eq: isometrization reactions}.

Let us start with a target that we wish the final tensor solution $\bfp_k$ to be accurate to order $10^{-4}$ in approximating the exact Poisson distribution \eqref{eq: isomerization Poisson}. 
This means that we should pick simulation parameters such that all sources of errors are kept below $10^{-4}$.
We choose the system volume $\Vol=500$ to be consistent with the simulations of the death-birth process \eqref{eq: birth-death reactions}. 
From Fig. \ref{fig: birth-death}(a), the modelling error at $\Vol=500$ would be far below $10^{-4}$.
The computational domain is chosen as $\D = \prod_{i=1}^{50} \D_i = [246,754]^{50}$, and accordingly to Fig. \ref{fig: birth-death}(b), the domain size $| \D_i |= 508$ keeps the artificial boundary error below $10^{-4}$.
We choose the equidistant grid size $h=4$ for all 50 dimensions, to keep the discretization error below $10^{-4}$ as suggested by Fig. \ref{fig: birth-death}(c).
We choose the tolerance of tensor rounding procedure to be $10^{-10}$ such that the tensor rounding error in Fig. \ref{fig: birth-death}(d) is not significant.
The tolerance for the tensor linear solver, AMEN, in the inverse iterations is chosen to be $10^{-3}$ to keep the algebraic error in Fig. \ref{fig: birth-death}(d) below $10^{-4}$.

\begin{figure}[t] 
\noindent
\rule{0pt}{0pt} 
\raise 5.2cm \hbox{(a)}
\hskip -7mm
\includegraphics[scale=0.34]{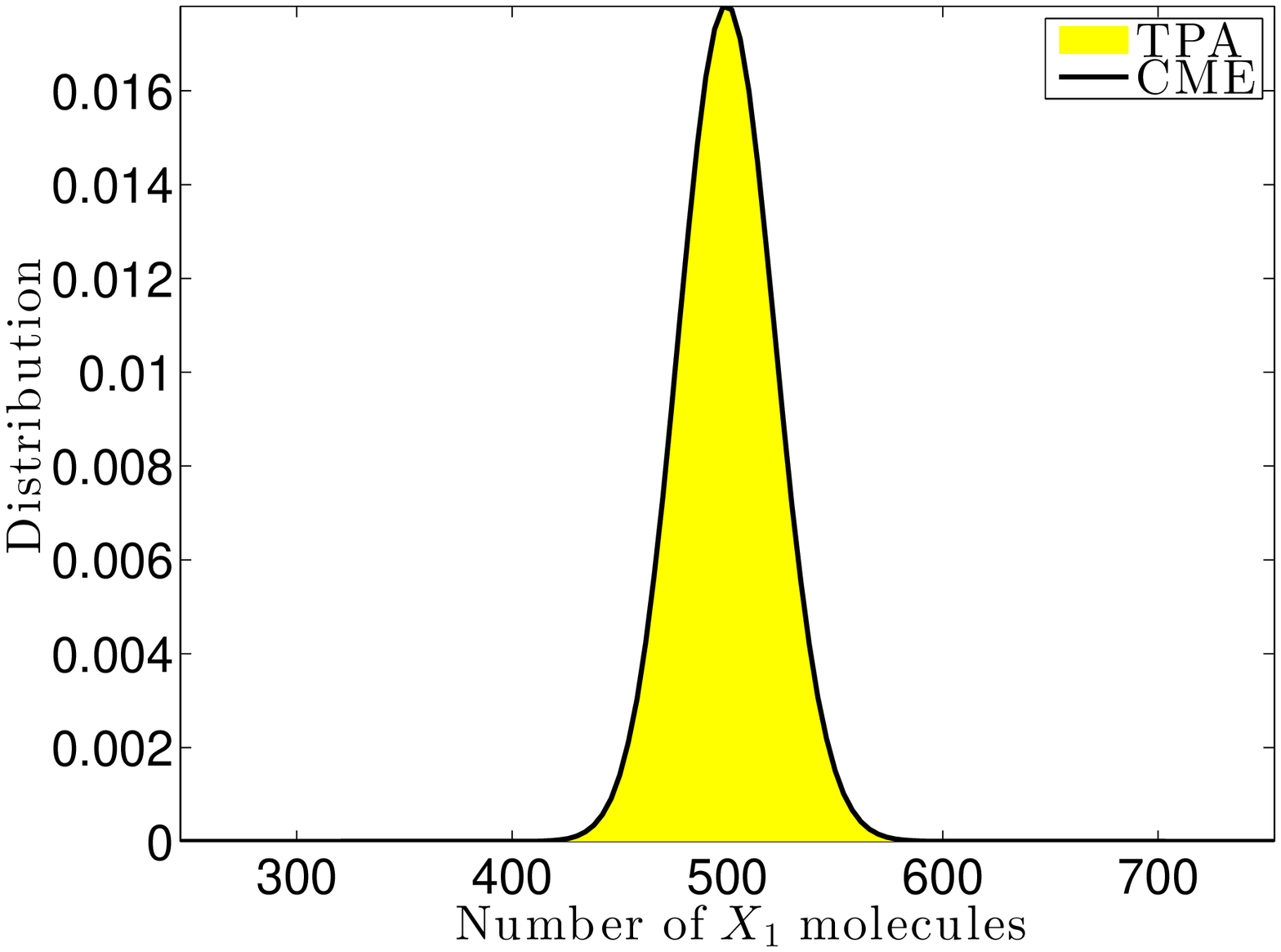}
\hskip 0mm
\raise 5.2cm \hbox{(b)}
\hskip -6mm
\includegraphics[scale=0.34]{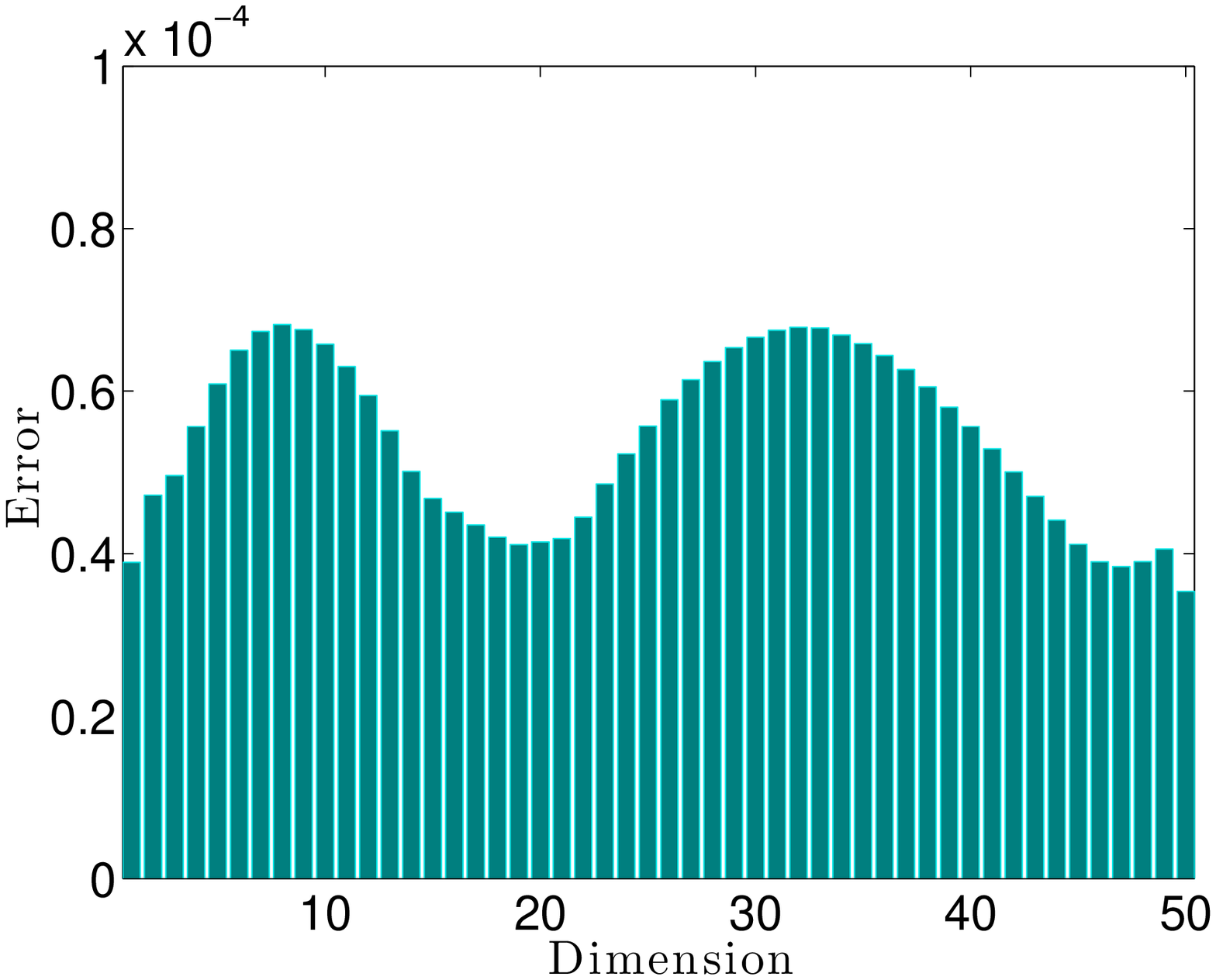}

\vskip 0.5mm

\hskip -0mm
\raise 5.2cm \hbox{(c)}
\hskip -7mm
\includegraphics[scale=0.34]{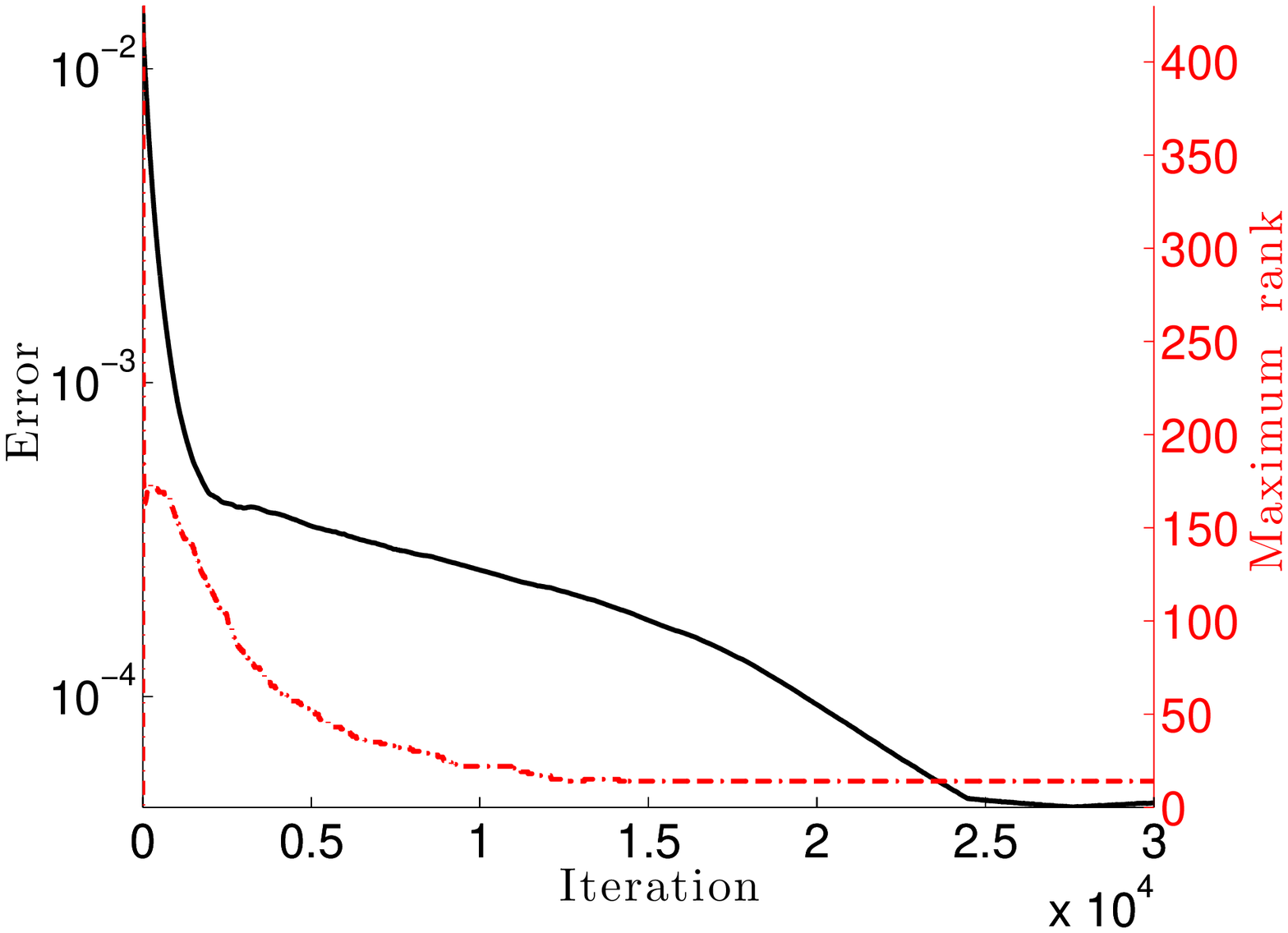}
\hskip 0mm
\raise 5.2cm \hbox{(d)}
\hskip -7mm
\includegraphics[scale=0.34]{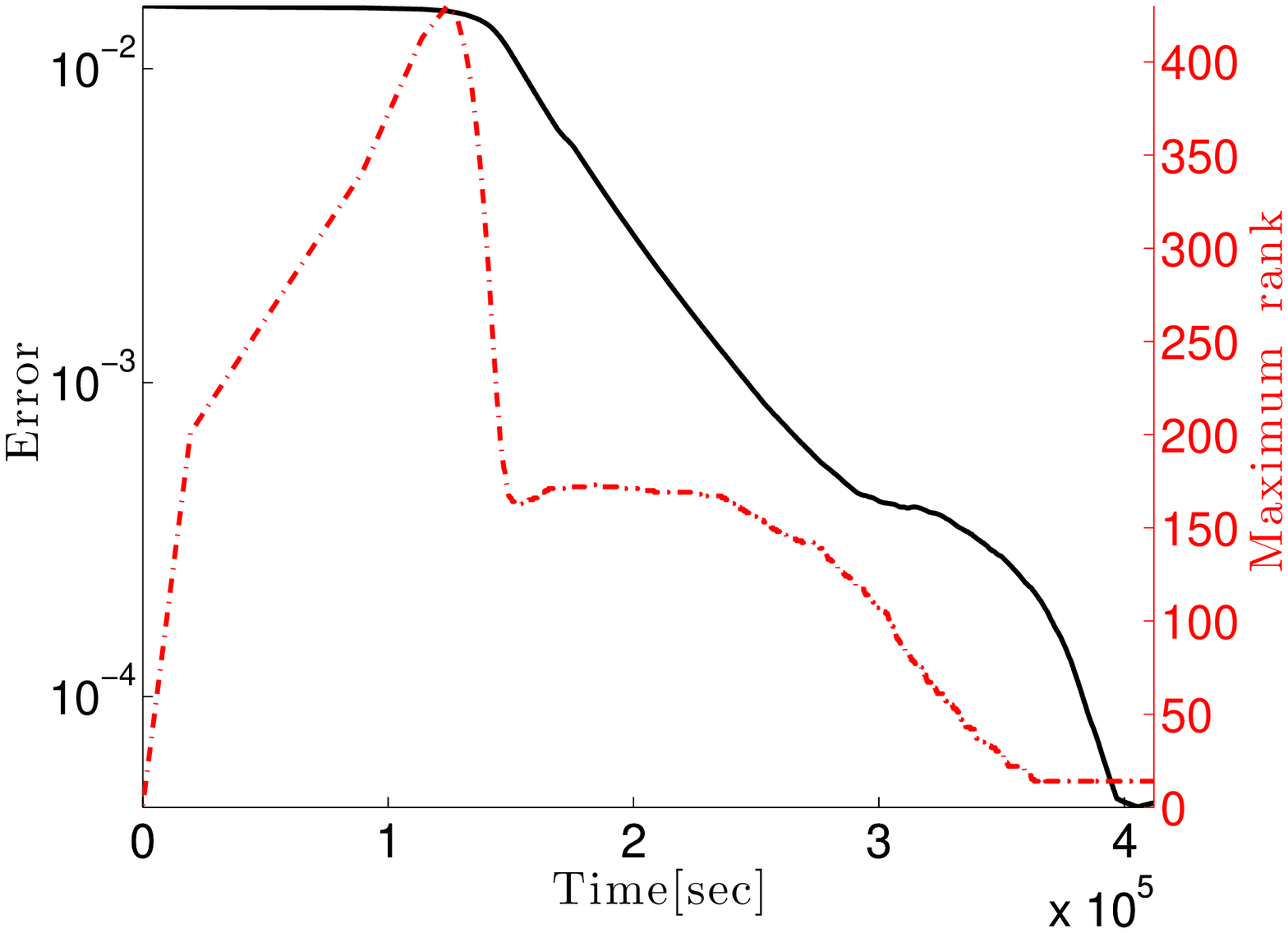}
\caption{ \label{fig: isomerization}
Steady state simulation results of the $50$-dimensional isomerization reaction chain \eqref{eq: isometrization reactions} in QTT format.
Parameters: reaction rates $k_j=1$ for $j \in [1,102]$, volume $\Vol = 500$, domain $\D = [246,754]^{50}$, grid size $h=4$, tensor rounding tolerance $\varepsilon_t = 10^{-10}$, shift value $\sigma=40$, and tolerance for tensor linear solver $\varepsilon_k = 10^{-3}$.
Storage requirement of tensor matrix $\mathbf{A}_t$ is $176144$ with maximum QTT rank $R_{\max}=13$, and the final tensor solution $\bfp_k$ is $61848$ with $R_{\max}=14$.
{\rm (a)} 
The computed marginal distribution for $X_1$ species versus the Poisson distribution $\Pms^1$ in \eqref{eq: isomerization Poisson} as the exact solution of the CME \eqref{eq: CME}.
{\rm (b)}
Error in $\ell^\infty$-norm of the marginal distributions in all 50 dimensions.
{\rm (c)}
Error convergence and maximum QTT rank against the number of inverse iterations.
Error is measured in $\ell^\infty$-norm between the marginal distribution of $X_1$.
{\rm (d)}
Error convergence and maximum QTT rank against the computational time.
}
\end{figure}

The simulation results and performances of the TPA are demonstrated in Fig. \ref{fig: isomerization}.
The computed tensor data agrees well with the exact Poisson distribution (see Fig. \ref{fig: isomerization}(a)), and the accuracy meets the pre-set target $10^{-4}$ in all 50 dimensions (see Fig. \ref{fig: isomerization}(b)).
The error convergence in Fig. \ref{fig: isomerization}(c) matches the prediction of Theorem \ref{theorem: algebiraic error convergence} that the error first decreases monotonically and then the convergence comes to a halt due to the inexactness in solving the tensor linear systems.
But we also observe the maximum QTT rank increases dramatically in the initial inverse iterations.
Larger tensor ranks give rise to quadratic complexity of basic arithmetic, and as a consequence, the first few iterations with large QTT ranks cost almost half of the computational time, while their effect in error convergence is not obvious at all (see Fig. \ref{fig: isomerization}(d)).
This refers to a major problem in the low-rank tensor computations that, even the final solution admits low-rank approximations, the growth in the ranks of the intermediate solutions might still kills the simulation.
This still remains as an open problem in this area.

Our 50-dimensional solution in QTT format has storage requirement to be $61848$, whereas this number would be $2.3\times 10^{105}$ if stored in a standard vector.
This demonstrate the effectiveness of the tensor approach for analysing high-dimensional stochastic models of GRNs. 
It is also worth pointing out that the total computational time is as long as $4 \times 10^5$ seconds using a personal laptop. 
Such heavy simulation highlights the crucial role of the error analysis as we have presented in the current paper, because such understanding not only has informed us about the accuracy of the method, but has also guided us with feasible choices of simulations parameters such that unnecessary repetitions could be avoided.

\section{Discussion}

In this paper we have presented a detailed mathematical and numerical study of the difference sources of errors in the recently proposed tensor approach \cite{liao2014parameter} for simulating high-dimensional stochastic models of GRNs.
The five sources of errors include: 
modelling error due to approximate the CME by a Fokker-Planck-type diffusion process; 
artificial boundary error due to the truncation of the infinite domain of definition into a computable bounded domain;
discretization error in the finite difference approximations;
tensor round error due to the tensor rounding procedure;
and algebraic error caused by the tensor-structured inverse power method.

These errors are like the stepping stones that bridge the gap between what we get and what we expect. 
We emphasise in this work that the total error of the TPA, as well as many other simulation methods in the literature, never solely relies on any particular source of errors, but is orchestrated by all of them.
On one hand, it warns us with the complication in the choice of simulation parameters to reduce the overall errors. 
On the other hand, it hints the flexibility to make computational trade-offs through controlling individual sources of errors, especially in high-dimensional computations.
As we posed in the birth-death example of section 7.1, if the system volume $\Vol$ has induced modelling error of order $\mathcal{O}(10^{-4})$, it is rarely necessary to pick a small grid size $h \ll 10$ or choose a small tolerance for the tensor linear solver $\varepsilon \ll 10^{-2}$, because it would not reduce the overall error that has already been significantly contributed by the modelling error (see Figs. \ref{fig: birth-death} and \ref{fig: birth-death algebraic}).
It is therefore reasonable to relax the restrictions on $h$ and $\varepsilon$, and the computational efficiency could be improved.

Notwithstanding that the presented analysis has been mainly tailored for the TPA, many results may be useful in a wider range of methods and applications.
The modelling error estimate in Theorem \ref{theorem: modellingerrdetailbalance} could be applied to analyse other stochastic simulation methods based on the Fokker-Planck formulations~\cite{gillespie2000the, sjoberg2009fokker, cotter2011a, cotter2013adaptive}.
The monotone difference scheme described in section 4.1 introduces the idea of applying different stencils to the positive and negative summands of the diffusion coefficients, while the existing schemes mainly separate the positive and negative parts~\cite{samarskii2002monotone}. 
For many elliptic and parabolic problems, the coefficients may be splitted into the summands that admit a separable form, then using our modified difference scheme, these problems can be directly equipped with tensor representations and solved in higher dimensions.
Finally, our study on the algebraic error could be applied to legitimise and analyse the use of inverse power method with tensor rank truncations in many other high-dimensional eigenvalue problems~\cite{beylkin2002numerical, hackbusch2012use}.

\section*{Acknowledgment}

The author would like to thank Professor Radek Erban and Dr Tom\'{a}\v{s} Vejchodsk\'{y}
for all of their careful, constructive and insightful comments that greatly contributed to improving the final version of the paper.
The research leading to these results has received funding from
the European Research Council under the European Community's
Seventh Framework Programme (FP7/2007–2013)/ERC grant
agreement no. 239870.



\bibliographystyle{unsrt}

\begin{thebibliography}{1}

\bibitem{samarskii2002monotone}
{\sc A. A. Samarskii, P. P. Matus, V. I. Mazhukin, and I. E. Mozolevski},
{ \em Monotone difference schemes for equations with mixed derivatives},
Comput. Math. Appl., 44 (2002), pp.~501--510.

\bibitem{carasso1969finite}
{\sc A. Carasso}, 
{\em Finite-difference methods and the eigenvalue problem for nonselfadjoint Sturm-Liouville operators}, 
Math. Comp., 23 (1969), pp.~717--729.



\bibitem{deif1995rigorous} {\sc A. S. Deif}, {\em Rigorous perturbation bounds for eigenvalues and eigenvectors of a matrix}, J. Comput. Appl. Math., 57 (1995), pp.~403--412.

\bibitem{munsky2006the}
{\sc B. Munsky, and M. Khammash},
{\em The finite state projection algorithm for the solution of the chemical master equation}, 
J. Chem. Phys., 124  (2006), pp.~044104.


\bibitem{gillespie1977exact} {\sc D. T. Gillespie}, {\em Exact Stochastic Simulation of Coupled Chemical Reactions}, J. Chem. Phys., 81 (1977), pp.~2340–-2361.

\bibitem{gillespie2000the}
{\sc D. T. Gillespie},
{\em The chemical Langevin equation},
J. Chem. Phys., 113 (2000), pp~297--306.

\bibitem{hitchcock1927the}
{\sc F. L. Hitchcock},
{\em The expression of a tensor or a polyadic as a sum of products},
J. Math. Phys., 6 (1927), pp.~164--189.

\bibitem{beylkin2002numerical}
{\sc G. Beylkin, and M. J. Mohlenkamp}, 
{\em Numerical operator calculus in higher dimensions}, 
Proc. Natl. Acad. Sci. USA, 99 (2002), pp.~10246--10251.

\bibitem{golub2000inexact} {\sc G. Golub and Q. Ye}, {\em Inexact inverse iteration for generalized eigenvalue problems},
BIT Numerical Mathematics, 40  (2000), pp.~671--684.

\bibitem{kramers1940brownian}
{\sc H. A. Kramers}, 
{\em Brownian motion in a field of force and the diffusion model of chemical reactions},
Physica 7 (1940), pp~284--304.

\bibitem{keller1965on}
{\sc H. B. Keller},
{\em On the accuracy of finite difference approximations to the eigenvalues of differential and integral operators},
Numer. Math., 7 (1965), pp.~412--419.


\bibitem{berestycki2014generalizations}
{\sc H. Berestycki and L. Rossi},
{\em Generalizations and Properties of the Principal Eigenvalue of Elliptic Operators in Unbounded Domains},
Comm. Pure Appl. Math., (2014). doi: 10.1002/cpa.21536



\bibitem{oseledets2011tensor} {\sc I.~V. Oseledets}, {\em Tensor-train decomposition}, SIAM J. Sci. Comp., 33
(2011), pp.~2295--2317.

\bibitem{oseledets2009breaking} {\sc I.~V. Oseledets and E.~E. Tyrtyshnikov}, {\em Breaking the curse of dimensionality, or how to use svd in many dimensions}, SIAM J. Sci. Comp., 31
(2009), pp.~3744--3759.

\bibitem{oseledets2011tensorb}
{\sc I. V. Oseledets, E. E. Tyrtyshnikov, and N. Zamarashkin}, 
{\em Tensor-train ranks for matrices and their inverses}, 
Comput. Methods Appl. Math., 11 (2011), pp.~394--403.

\bibitem{rybak2004monotone}
{\sc I. V. Rybak},
{\em Monotone and conservative difference schemes for elliptic equations with mixed derivatives},
Math. Model. Anal., 9 (2004), pp.~169--178.

\bibitem{moyal1949the}
{\sc J. E. Moyal},
{\em The distribution of wars in time},
J. R. Stat. Soc., 11 (1949), pp.~446--449.

\bibitem{gary1965computing} 
{\sc J. Gary},
{\em Computing eigenvalues of ordinary differential equations by finite differences},
Math. Comp., 19 (1965), pp.~365--379.

\bibitem{kuttler1970finite}
{\sc J. R. Kuttler},
{\em Finite difference approximations for eigenvalues of uniformly elliptic operators}, 
SIAM J. Numer. Anal., 7 (1970), pp.~206--232.


\bibitem{grasedyck2010hierarchical}
{\sc L. Grasedyck}, 
{\em Hierarchical singular value decomposition of tensors}, 
SIAM. J. Matrix Anal. \& Appl.,  31 (2010), pp.2029--2054.

\bibitem{tucker1966some}
{\sc L. R. Tucker}, 
{\em Some mathematical notes on three-mode factor analysis}, 
Psychometrika, 31 (1966), pp.~279--311.

\bibitem{gibson2000efficient} 
{\sc M. A. Gibson and J. Bruck}, {\em Efficient exact stochastic simulation 
of chemical systems with many species and many channels}, 
J. Phys. Chem. A, 104 (2000), pp~1876--1889.


\bibitem{vankampen1992stochastic}
{\sc N. G. Van Kampen}
{\em Stochastic processes in physics and chemistry},
North-Holland, Amsterdam,1992.

\bibitem{halmos1974finite} 
{\sc P. Halmos}, {\em Finite dimensional vector spaces}, Springer-Verlag, New York, 1974.

\bibitem{matus2004difference}
{\sc P. Matus, and I. Rybak},
{\em Difference schemes for elliptic equations with mixed derivatives},
Comput. Methods Appl. Math., 4 (2004), pp.~494--505.

\bibitem{sjoberg2009fokker}
{\sc P. Sjoberg, P. Lotstedt, and J. Elf},
{\em Fokker–Planck approximation of the master equation in molecular biology}, 
Comput. Visual. Sci., 12 (2009), pp.~37--50.

\bibitem{bellman1961adaptive}
{\sc R. Bellman},
{\em Adaptive control processes: a guided tour},
Princeton university press, Princeton, 1961.


\bibitem{grima2011how} 
{\sc R. Grima, P. Thomas, and A. V. Straube}, 
{\em How accurate are the nonlinear chemical Fokker-Planck and chemical Langevin equations?}, J. Chem. Phys., 135 (2011), pp~084103.


\bibitem{cotter2013adaptive}
{\sc S. Cotter, T. Vejchodsky, and R. Erban}, 
{\em Adaptive finite element method assisted by stochastic simulation of chemical systems},
SIAM J. Sci. Comput., 35 (2013), pp.~B107--B131.

\bibitem{cotter2011a}
{\sc S. Cotter, K. Zygalakis, I. Kevrekidis, and R. Erban},
{\em A constrained approach to multiscale stochastic simulation of chemically reacting systems}, 
J. Chem. Phys., 135 (2011), p.~094102.

\bibitem{holtz2012the} {\sc S. Holtz, T. Rohwedder, and R. Schneider}, 
{\em The alternating linear scheme for tensor optimization in the tensor train format}, SIAM J. Sci. Comp., 34
(2012), pp.~A683--A713.

\bibitem{liao2016high} {\sc S. Liao}, 
{\em High-dimensional problems in stochastic modelling of biological processes}, PhD thesis, University of Oxford (2016).


\bibitem{liao2014parameter} {\sc S. Liao, T. Vejchodsky, and R. Erban}, {\em Tensor methods for parameter estimation and bifurcation analysis of stochastic reaction networks}, J. R. Soc. Interface., 12 (2015).

\bibitem{dolgove2014simultaneous} {\sc S. V. Dolgov, and B. N. Khoromskij}, 
{\em Simultaneous state-time approximation of the chemical master equation using tensor product formats}, 
Numer. Linear Algebra Appl.,
22 (2014), pp.~197--219.

\bibitem{dolgov2012fast} {\sc S. V. Dolgov, B. N. Khoromskij, and I. V. Oseledets}, {\em Fast solution of parabolic problems in the tensor train/quantized tensor train format with initial application to the Fokker--Planck Equation}, SIAM J. Sci. Comp., 34 (2012), pp.~A3016--A3038.

\bibitem{dolgov2013alternating1} {\sc S. V. Dolgov and D. V. Savostyanov}, {\em Alternating minimal einergy methods for linear systems in higher dimensions. part I: SPD systems}, arXiv preprint arXiv:1301.6068, (2013).

\bibitem{dolgov2013alternating2} {\sc S. V. Dolgov and D. V. Savostyanov}, {\em Alternating minimal einergy methods for linear systems in higher dimensions. part II: faster algorithm and application to nonsymmetric systems}, arXiv preprint arXiv:1304.1222, (2013).

\bibitem{kolda2009tensor}
{\sc T. G. Kolda and B. W. Bader},
{\em Tensor decompositions and applications}, 
SIAM review, 51 (2009), pp.~455--500.

\bibitem{jahnke2008a} 
{\sc T. Jahnke and W. Huisinga}, 
{\em A dynamical low-rank approach to the chemical master equation},
B. Math. Biol., 70 (2008), pp.~2283--2302.

\bibitem{jahnke2007solving} {\sc T. Jahnke and W. Huisinga},
{\em Solving the chemical master equation for monomolecular reaction systems analytically}, 
J. Math. Biol., 54 (2007), pp.~1--26.

\bibitem{kurtz1978the} {\sc T. G. Kurtz},
{\em Strong approximation theorems for density dependent Markov chains},
Stoch. Proc. Appl., 6 (1978), pp.~223--240.

\bibitem{kazeev2014direct} {\sc V.~A. Kazeev, M. Khammash, M. Nip, and C. Schwab}, 
{\em Direct solution of the chemical master equation using quantized tensor trains}, PLoS Comp. Biol., 10
(2014), pp.~e1003359.

\bibitem{kazeev2012low} {\sc V. A. Kazeev and B. N. Khoromskij}, 
{\em Low-rank explicit QTT representation of the Laplace operator and its inverse}, SIAM J. Matrix Anal. Appl., 33 (2012), pp~742--758.


\bibitem{hackbusch2005hierarchical} {\sc W. Hackbusch, B. N. Khoromskij, and E. E. Tyrtyshnikov}, {\em Hierarchical Kronecker tensor-product approximations}, J. Numer. Math. 13 (2005), pp.~119--156.

\bibitem{hackbusch2012use}
{\sc W. Hackbusch, B. N. Khoromskij, S. Sauter, and E. E. Tyrtyshnikov},
{\em Use of tensor formats in elliptic eigenvalue problems},
Numer. Linear Algebra Appl., 19 (2012), pp.~133--151.


\bibitem{feng2013recent}
{\sc X. Feng, R. Glowinski, and M. Neilan}, 
{\em Recent developments in numerical methods for fully nonlinear second order partial differential equations}, 
SIAM Review, 55 (2013), pp.~205--267.

\bibitem{cao2004the} {\sc Y.  Cao, D. T. Gillespie, and L. R. Petzold}, 
{\em The slow-scale stochastic simulation algorithm}. 
J. Chem. Phy., 122 (2004), pp.~014116.



\end{thebibliography}

\end{document}